%% file: main.tex
\documentclass[letterpaper,twocolumn,10pt]{article}
\usepackage{usenix-2020-09}

\usepackage[utf8]{inputenc}
\usepackage{hyperref}
\usepackage{listings}
\usepackage{color,xcolor}
\usepackage{multirow}
\usepackage{sverb}
\usepackage{graphicx}
\usepackage{soul}
\usepackage{subfig}
\usepackage[norndcorners,customcolors]{hf-tikz}
\hfsetbordercolor{yellow}
\hfsetfillcolor{yellow}

\usepackage{verbatim}

\usepackage{tablefootnote}

\usepackage{threeparttable}

\usepackage[T1]{fontenc}
\usepackage{lmodern}
\usepackage[ngerman,english]{babel}
\usepackage{amsmath}
\usepackage{mathtools}

\usepackage{graphics}

\usepackage[ruled, vlined, linesnumbered]{algorithm2e}
\usepackage{makecell} 

\usepackage{pifont}
\newcommand{\cmark}{\text{\ding{51}}}%
\newcommand{\xmark}{\text{\ding{55}}}%

\usepackage{flexisym} 

\usepackage{listings}

\usepackage{float}
\usepackage{mwe}

\usepackage{alltt}

\definecolor{mGreen}{rgb}{0,0.6,0}
\definecolor{mGray}{rgb}{0.5,0.5,0.5}
\definecolor{mPurple}{rgb}{0.58,0,0.82}
\definecolor{backgroundColour}{rgb}{0.95,0.95,0.92}

\lstdefinestyle{CStyle}{
    backgroundcolor=\color{white},   
    commentstyle=\color{mGreen},
    keywordstyle=\color{magenta},
    numberstyle=\tiny\color{mGray},
    stringstyle=\color{black},
    basicstyle=\fontsize{6.5}{11}\selectfont\ttfamily,
    breakatwhitespace=false,         
    breaklines=true,                 
    captionpos=b,                    
    keepspaces=true,                 
    numbers=left, 
    literate = {-}{-}1,
    keywordstyle=\color{blue}, 
    numbersep=5pt,                  
    showspaces=false,                
    showstringspaces=false,
    showtabs=false,     
    tabsize=2,
    language=C
}
\newcommand{\lmcas}{\mbox{\textsc{LMCAS}}\xspace}

\newcommand{\ma}[1]{\textbf{\color{orange}\emph{M: #1}}}

\newcommand{\twr}[1]{\textbf{{\color{blue}\emph{T: #1}}}}

\newcommand{\revUN}[1]{#1}

\newcommand{\rev}[1]{#1}

\definecolor{bottlegreen}{rgb}{0.0,0.42,0.31}

\newcommand{\vr}[1]{\textbf{{\color{red}\emph{V: #1}}}}
\newcommand{\Omit}[1]{}

\colorlet{shadecolor}{gray!40}

\usepackage[normalem]{ulem} 
\usepackage{pxfonts} 

\usepackage{enumitem}
\setlist{nosep,leftmargin=\parindent}

\usepackage{times}

\title{Lightweight, Multi-Stage, \\ Compiler-Assisted Application Specialization}

\author{\vspace{-0.5cm}\\
  Mohannad Alhanahnah,
  Rithik Jain,
  Vaibhav Rastogi,
  Somesh Jha,
  Thomas Reps \\[3pt]
  {\rm UW--Madison}\\[1cm]
}

\begin{document}

\maketitle
\input{abstract}

\input{introduction}
\input{motivation}

\input{approach_V2}
\input{implementation}
\input{evaluation_usenix}

\input{discussion_V2}

\input{relatedWork}

\input{conclusion}

\bibliographystyle{abbrv}
\bibliography{ref.bib}

\input{appendix}

\end{document}

%% file: abstract.tex
\vspace{-0.3cm}
\begin{abstract}
Program debloating aims to enhance the performance and reduce the attack surface of bloated applications.
Several techniques have been recently proposed to \revUN{specialize programs}.
These approaches are either based on unsound strategies or demanding techniques, leading to unsafe results or a high-overhead debloating process.
In this paper,
\revUN{we address these limitations by applying partial-evaluation principles to generate specialized applications. Our approach relies on a simple observation that an application typically consists of configuration logic, followed by the main logic of the program.
The configuration logic specifies what functionality in the main logic should be executed. \lmcas performs partial interpretation to capture a precise program state of the configuration logic based on the supplied inputs. \lmcas then applies partial-evaluation optimizations to generate a specialized program by propagating the constants in the captured partial state, eliminating unwanted code, and preserving the desired functionalities.}

Our evaluation of \lmcas ---on commonly used benchmarks and real-world applications--- shows that it successfully removes unwanted features while preserving the functionality and robustness of the deblated programs, runs faster than prior tools, and reduces the attack surface of specialized programs. 
\lmcas runs $1500$x, $4.6$x, and $1.2$x faster than the state-of-the-art debloating tools CHISEL, RAZOR, and OCCAM, respectively;
achieves $25\%$ reduction in the binary size;
reduces the attack surface of code-reuse attacks by removing $51.7\%$ of the total gadgets and eliminating 83\% of known CVE vulnerabilities.

\end{abstract}

%% file: introduction.tex
\section{Introduction} \label{sec:introduction}
The software stack is becoming increasingly bloated. This software growth decreases performance and increases security vulnerabilities.
\emph{Software debloating} is a mitigation approach that downsizes programs while retaining certain desired functionality.
Although static program debloating can thwart unknown possibilities for attack by reducing the attack surface~\cite{attackSurface},
prior work has generally not been effective
, due to the 
overapproximation of static program analysis:
a lot of bloated code remains because these tools statically determine the set of functions to be removed using a combination of analysis techniques, such as unreachable-function analysis and global constant propagation~\cite{BlankIt}.
More aggressive debloating approaches (e.g., RAZOR~\cite{RAZOR} and Chisel~\cite{Chisel}) can achieve more reduction;
however, \revUN{they involve demanding techniques:} the user needs to define a comprehensive set of test cases to cover the desired functionalities, \revUN{generate many traces of the program, and perform extensive instrumentation}.
\revUN{The computational expense of these steps leads to a high-overhead debloating process.
The work on RAZOR acknowledges the challenge of generating test cases to cover all code, and incorporates heuristics to address this challenge.} Furthermore, these aggressive approaches often break soundness, which can lead the debloated programs to crash or execute incorrectly.
These issues make such approaches unsafe and impractical~\cite{BlankIt}.


Partial evaluation is a promising program-specialization technique.
It is applied in prior work~\cite{OCCAM,TRIMMER}, but prior work has suffered from the overapproximation inherent in static program analysis.
In particular, existing implementations rely only on the command-line arguments to drive the propagation of constants.
However, they do not capture precisely the set of variables that are affected by the supplied inputs.
Constant propagation is performed only for global variables that have one of the base types (\texttt{int}, \texttt{char}), and no attempt is made to handle compound datatypes (\texttt{struct}). Therefore, 
this approach leaves a substantial amount of unwanted code in the ``debloated program'', which 
reduces the security benefits and does not preserve the behaviour of the program after debloating due to unsound transformations (i.e., incorrect constant folding~\cite{TRIMMER}). (our evaluation in~\S\ref{sec:evaluation} shows multiple programs debloated by OCCAM are crashing or behave unexpectedly.)
\Omit{\vr{Only one point, so why (i)?\ma{thanks}}}
\Omit{\twr{I don't understand the second point, which seems backwards: it will \emph{include} the features that are controlled by the variables that partial evaluation ignores.\Omit{\ma{to address John's concern about soundness mentioned in the next statement. So PE in prior work can also exclude features, I had this case when OCCAM returns empty output}}}
\twr{You are thinking about a point that is far to complicated for the introduction.  If you are worried about the word "soundness", then remove the word, remove the label "(i)" above, remove all of point "(ii)" below, and end the previous sentence at "reduces the security benefits".}}\Omit{\ma{(ii) can exclude required features that are enabled by the ignored variables.}}
\Omit{\vr{Incomplete sentence. Also, we do not have an evaluation of performance.}\ma{thanks}}

In this paper, we present Lightweight Multi-Stage Compiler-Assisted Application Specialization (\lmcas), a new software-debloating framework.
\lmcas relies on the observation that, in general, programs consist of two components:
(a) \textit{configuration logic}, in which the inputs are parsed, and
(b) \textit{main logic}, which implements the set of functionalities provided by the program.
We call the boundary between the two divisions the \textit{neck}. 
\revUN{\lmcas captures a partial state of the program by interpreting the configuration logic based on the supplied inputs. The partial state comprises concrete values of the variables that are influenced by the supplied inputs, 
\lmcas then applies partial-evaluation optimizations to generate the specialized program. These optimizations involve: converting the influenced variables at the neck into constants, applying constant propagation, and performing multiple stages of standard and customized compiler optimizations. 
}

\revUN{\lmcas makes significant and novel extensions to make debloating much safer in a modern context.
The extensions involve optimizing the debloating process and improving its soundness.
Specifically, we optimize the debloating process by introducing the \textit{neck} concept, which eliminates demanding techniques that require: (1) executing the whole program, (2) generating many traces, (3) performing extensive instrumentation, and (4) obtaining a large set of tests.
We demonstrate the soundness of our approach by validating the functionality of $23$ programs after debloating and under various settings.
The achieved soundness is driven by: capturing a precise partial state of the program, supporting various data types, and performing guided constant conversion and clean-up.
}
Our evaluation demonstrates that \lmcas is quite effective: on average, \lmcas achieves $25\%$ reduction in the binary size and $40\%$ reduction in the number of functions in the specialized applications.
\lmcas reduces the attack surface of code-reuse attacks by removing $51.7\%$ of the total gadgets, and eliminates $83\%$ of known CVE vulnerabilities.
On average, \lmcas runs $1500$x, $4.6$x, and $1.2$x faster than the state-of-the-art debloating tools CHISEL, RAZOR, and OCCAM, respectively.
\revUN{Hence, LMCAS strikes a favorable trade-off between functionality, performance, and security.
}

The contributions of our work are as follows:
\begin{enumerate}
    \item \revUN{We propose the novel idea of dividing programs into configuration logic and main logic to reduce the overhead of debloating process.
    }
    \item \revUN{We develop a neck miner to enable the identification of such boundary with high accuracy, substantially alleviating the amount of manual effort required to identify the neck.}
    \Omit{\twr{The previous sentence is way off the mark.
    It makes it sound like the paper is about dynamic analysis, and that partitioning is a technique that you have introduced to improve a dynamic-analysis tool.}}
    \item \revUN{
    We apply the principles of partial evaluation to generate specialized programs based on supplied inputs. A partial program state is captured by conducting partial interpretation for the program by executing the configuration logic according to the supplied inputs. The program state is enforced by applying compiler optimizations for the main logic to generate the specialized program.}
    \item We develop the \lmcas prototype based on LLVM. \lmcas harnesses symbolic execution to perform partial interpretation and applies a set of LLVM passes to perform the compiler optimizations. We also carried out an extensive evaluation based on real-world applications to demonstrate the low overhead, size reduction, security, and practicality of \lmcas.
    \Omit{\twr{I would split this point in two: the ``principles'' and the prototype.
    The principles are the core idea found in the paper, and should go either first or second in the list of contributions (probably second, after the idea that many programs can be split into configuration and main logic. The prototype can be folded into the bullet point about evaluation.}}
    \item We will make \lmcas implementation 
    and its artifacts available to the community.
\end{enumerate}

%% file: motivation.tex
\section{Motivation and Background} \label{sec:motivation}

In this section, we present a motivating example, and review necessary background material on program analysis and software debloating used in the remainder of the paper.
Finally, we discuss debloating challenges illustrated by the motivating example, and describe our solutions for addressing these challenges.
\subsection{Motivating Example}
Listing~\ref{wc} presents a scaled-down version of the UNIX word-count utility \texttt{wc}.
It reads a line from the specified stream (i.e., \texttt{stdin}), counts the number of lines and/or characters in the processed stream, and prints the results.
Thus, this program implements the same action that would be obtained by invoking the UNIX word-count utility with the \texttt{-lc} flag (i.e., \texttt{wc -lc}). Although Listing~\ref{wc} merely supports two counting options, it is still bloated if the user is only interested in enabling the functionality that counts the number of lines.
\input{listings/toy_wc_struct}
Put another way, Listing~\ref{wc} goes
against the \textit{principle of least privilege}~\cite{plp}~\cite{Cimplifier}:
code that implements unnecessary functionality can contain security vulnerabilities that an attacker can use to obtain control or deny service---bloated code may represent an opportunity for privilege escalation.
For instance, the character-count functionality ``\texttt{wc -c}'' processes and decodes the provided stream of characters via the function \texttt{decodeChar} (Line 17 of Listing~\ref{wc}).
An attacker might force it to process special characters that \texttt{decodeChar} cannot handle~\cite{wc-bug}.
More broadly, attackers can supply malicious code as specially crafted text that injects shellcode and thereby bypasses input restrictions~\cite{EnglishShellcode}.
\Omit{\vr{Since shellcode here is English text, the characters will be easily
(correctly) processed by wc, I do not see how supplying a file containing
shellcode will suddenly break wc.} \ma{I missed to mention, this is not normal English text, it's a crafted text of the malicious code. So this text can bypass input restrictions}}
Downsizing the program is a way to reduce its attack surface because the ``\texttt{wc -l}'' functionality does not require the character-processing code used in ``\texttt{wc -c}'':
the call to the function \texttt{decodeChar} is completely absent in the specialized version for ``\texttt{wc -l}''.
~\revUN{To achieve this goal, debloating should be performed safely without applying high-demand techniques. The next section describes our approach for handling these challenges.}

\subsection{Background}
\noindent Partial evaluation~\cite{partialEvaluation,slicing,tutorialPE} is an optimization and specialization technique that precomputes program expressions in terms of the known static input ---i.e., a subset of the input that is supplied ahead of the remainder of the input.
The result is another program, known as the \emph{residual program}, which is a specialization of the original program. To create the residual program, a partial evaluator performs optimizations such as loop unrolling, constant propagation and folding, function inlining, etc.~\cite{partialEvaluation}.

Partial evaluation and symbolic execution~\cite{SymbolicExecution} both generalize standard interpretation of programs, but there are key differences between them. Specifically, symbolic execution interprets a program using symbolic input values, while partial evaluation precomputes program expressions and simplifies code based on the supplied inputs.
Unlike partial evaluation, one result of symbolic execution is a set of expressions \Omit{\twr{"Expressions" is correct; formulas are a different kind of result that SE gives, for characterizing what input condition is required to follow a particular path.}}for the program's output variables. Because their capabilities are complementary, the two techniques have been employed together to improve the capabilities of a software-verification tool~\cite{Interleaving}.

\revUN{{LLVM~\cite{LLVM:CGO04}} provides robust compiler infrastructures for popular programming languages, including C and C++, and supports a wealth of compiler analyses and optimizations that make it suitable for developing new compiler transforms.
LLVM operates on its own low-level code representation known as the LLVM intermediate representation (LLVM IR).
LLVM is widely used in both academia and industry.
}

\subsection{Challenges and Solutions} \label{subsec:challenges}
In this section, we formalize the program-specialization problem illustrated in Listing~\ref{wc}. 
In general, there is (i) a program P that provides a set of functionalities $F$,
and (ii) an input space $I$ that contains a set of values that enable certain functionalities in $F$.
Typically, one or more functionalities are enabled based on a set of supplied inputs $I_s$, which are provided as part of a command-line argument or configuration file. 
Generating a specialized program $P\textprime$ based on the set of supplied inputs $I_s$ requires identifying a set of variables $V_s=\{v_0, v_1, .., v_n\}$ that are influenced by the supplied inputs, and a corresponding set of constant values $C_s=\{c_0, c_1, .., c_n\}$. The relationship between $V_s$ and $C_s$ is bijective and $I_s \subset I$. 
For generating a specialized program $P\textprime$ that (i) retains the required functionalities based on the supplied inputs $I_s$, and (ii) removes irrelevant functionalities, we need to address the challenges discussed below: 

\Omit{\twr{You seemn to have adopted the term ``dynamic analysis,'' which seems completely wrong to me: you are just executing the program along a single path, so in what sense is any ``analysis'' going on?
I thought we were calling it ``partial execution.''}
\twr{
Also, you also keep referring to ``the overhead of dynamic analysis'' in other work.  I don't think that that comparison helps in this section.
Let's stick to describing our work, and make comparisons only in the Introduction and Related Work.}}
\textbf{Challenge 1:} 
\revUN{how to \emph{optimize} the debloating process and avoid high-demand techniques}?

\textbf{Solution.} \revUN{To address this challenge, we propose to interpret the program partially up to a certain point, instead of executing the whole program. We can achieve this partial interpretation by relying on the observation that, in general, programs consist of two components: (a) \textit{configuration logic}, in which the inputs (from the input space $I$) are parsed, and (b) \textit{main logic}, which implements the set of functionalities $F$. We call the boundary point the \textit{neck}.
The \textit{partial interpreter} needs only part of the program state to be available by executing the program upto the neck.
By this means, we optimize the debloating process, yet obtain a precise characterization of the set of variables $V_s$ that are influenced by the supplied arguments.
We then convert these variables to constants, based on the constant values $C_s$ identified by the \textit{partial interpreter}.
These values are then propagated to other parts of the program via partial evaluation.

Consider Listing~\ref{wc} again.
The program $\texttt{wc}$ provides two functionalities, $F = \{\textit{counting\_lines}, \textit{counting\_characters}\}$ and these functionalities can be activated through the two inputs $I = \{l, c\}$. For generating the specialized program $P\textprime$ that retains the \emph{counting\_lines} functionality (i.e., ``\texttt{wc -l}'') based on the supplied input $I_s = \{l\}$, we interpret program $P$ upto the neck (i.e., Line $15$) to identify the set of influenced variables $V_s = \{\texttt{flag->count\_chars}$, $\texttt{flag->count\_lines}$, $\texttt{total\_lines}$, $\texttt{total\_chars}\}$ and the corresponding constant values $C_s = \{0,1,0,0\}$ (the partial state of $P$). We supply this information to the partial evaluator to generate the specialized program.}

\textbf{Challenge 2:}
\revUN{how to simplify the program \emph{sufficiently}, while ensuring that it operates \emph{correctly}, and preserve its functionality and soundness?}

\textbf{Solution.} \revUN{The combination of \textit{partial interpretation} followed by partial-evaluation optimizations holds the promise of achieving significant debloating.
To achieve this promise and preserve the program semantics, it is necessary to handle various data types and complex data structures (i.e., string, pointers, and structs).
By using a precise model of the programming language's semantics, more information about variables and their values is made available, which in turn enables more optimizations to be carried out during program specialization.
Therefore, we need to capture a broad spectrum of variables. For instance, the scaled-down word-count in Listing~\ref{wc} contains a stack variable (\texttt{flag}) and two global variables (\texttt{total\_lines} and \texttt{total\_chars}). Various data types need to be supported as well: the global variables are integers, whereas the stack variable \texttt{flag} is a pointer to a struct that consists of two fields (\texttt{count\_line} and \texttt{count\_chars}). Supporting these various types of variables provides \lmcas the capability to perform safe debloating and maintain soundness.} 

%% file: listings/toy_wc_struct.tex

\begin{lstlisting}[style=CStyle, caption={A scaled-down version of the \texttt{wc} utility. \rev{Highlighted statements are eliminated after debloating with ``\texttt{wc -l}''}},label={wc},escapechar=@]
struct Flags {
	char count_chars;
	int count_lines; };
int total_lines = 0;
@\colorbox{shadecolor}{int total_chars = 0;}@
int main(int argc, char** argv){
	struct Flags *flag;
	flag = malloc(sizeof(struct Flags));
	@\colorbox{shadecolor}{flag->count_chars = 0;}@
	@\colorbox{shadecolor}{flag->count_lines = 0;}@
	@\colorbox{shadecolor}{if (argc >= 2)\{}@
		@\colorbox{shadecolor}{for (int i = 1; i < argc; i++) \{}@ 
			@\colorbox{shadecolor}{if (!strcmp(argv[i], "-c")) flag->count_chars = 1;}@
			@\colorbox{shadecolor}{if (!strcmp(argv[i], "-l")) flag->count_lines = 1; \}\}}@
	char buffer[1024];
	while (fgets(buffer, 1024,stdin)){
		@\colorbox{shadecolor}{if (flag->count_chars) total_chars += decodeChar(buffer);}@
		@\colorbox{shadecolor}{if (flag->count_lines)}@ total_lines++;}
	@\colorbox{shadecolor}{if (flag->count_chars) printf("\#Chars = \%d", total_chars);}@
	@\colorbox{shadecolor}{if (flag->count_lines)}@ printf("#Lines = %d", total_lines); }
\end{lstlisting}

%% file: approach_V2.tex
\section{\lmcas Framework}
\label{sec:approach}

\begin{figure}[tb]
    \centering
    \includegraphics[width=9cm, height=2.5cm]{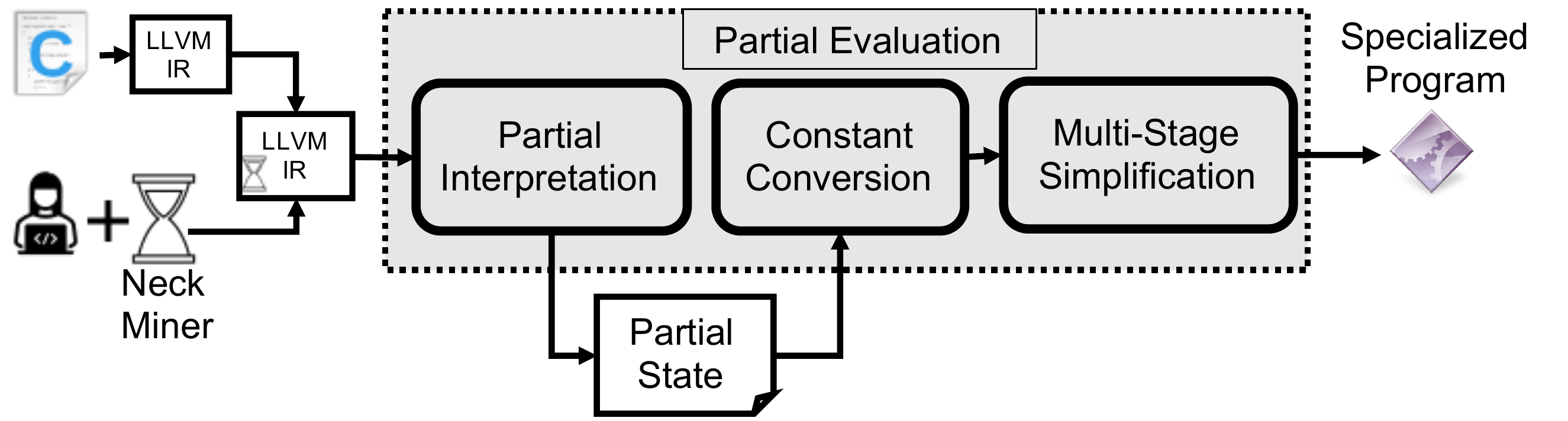}
    \caption{\revUN{\lmcas Workflow.}}
    \label{fig:approach}
\end{figure}

This section introduces \lmcas, a lightweight debloating framework that uses sound analysis techniques to generate specialized programs. 
Figure~\ref{fig:approach} illustrates the architecture of \lmcas. 
The debloating pipeline of \lmcas receives as input the 
whole-program LLVM bitcode of program $P$, and\Omit{\twr{wording changes}}
performs the following major phases to generate a specialized program $P\textprime$ as bitcode, which is ultimately converted into a binary executable.

\begin{itemize}
    \item \textbf{Neck Miner (Section~\ref{subsec:neckSpec}).} 
    \revUN{Receives the program to be specialized and modifies it by adding a special function call that marks the neck. Section~\ref{subsec:neckSpec} describes our approach to identifying the neck based on heuristic and structural analysis.} 
    
    \item \textbf{Partial Interpretation (Section~\ref{subsec:SE}).} 
    Interprets the program 
    up to the neck \revUN{(i.e., terminates after executing the special function call inserted by the neck miner)}, based on the supplied inputs that control which functionality should be supported by the specialized application.
    The output of this phase provides a precise partial state of the program at the neck.\revUN{ This partial state comprises the variables that have been initialized during the partial interpretation and their corresponding values.}
    
    \item \textbf{Constant Conversion (Section~\ref{subsec:cc}).}
    Incorporates into the program the partial state captured by the partial interpretation. 
    It converts the variables and their corresponding values captured in the partial state (i.e., $V_s$ and $C_s$) to settings of constants at the neck. This phase also provides the opportunity to boost the degree of subsequent optimization steps by supporting the conversion of multiple kinds of variables to constants.
    
    \item \textbf{Multi-Stage Simplification (Section~\ref{subsec:msp}).}
    Applies selected standard and customized LLVM passes for optimizing the program and removing unnecessary functionality.
    These optimization steps are arranged and tailored to take advantage of the values introduced by the constant-conversion phase. 
\end{itemize}

\subsection{Neck Miner} \label{subsec:neckSpec}
\input{neck}

\subsection{Partial Interpretation (PI)} \label{subsec:SE}
    Partial interpretation is a supporting phase whose goal is to identify---at the neck---the set of variables (and their values) that contribute to the desired functionality.
    Partial interpretation is performed by running a symbolic-execution engine, starting at program entry, and executing the program (using partial program states, starting with the supplied concrete values) along a single path to the neck.
    After partial interpretation terminates, the partial state is saved, and the values of all variables are extracted.
    Different types of variables are extracted, including base types (e.g., \texttt{int}, \texttt{char}) and constructed/compound types (e.g., \texttt{enum}, pointer, array, and \texttt{struct}). 
    
    \Omit{\twr{Omit starting with
    ``Our approach supports capturing these various types of data types to perform a sound analysis.''  I think it is something you've written to defend/distinguish your work w.r.t. CHISEL, OCCAM, and TRIMMER, but it is only intrusive in the presentation of \emph{your} work.} \ma{I agree}}
    
    \Omit{\twr{Does the following sentence capture what you intended to say?} \ma{ Yes}}
    Consider a network-monitoring tool, such as \texttt{tcpdump}, that supports multiple network interfaces, and takes as input a parameter that specifies which network interface to use.
    A specialization scenario might require monitoring a specific interface (e.g., Ethernet) and capturing a specific number of packets: for \texttt{tcpdump}, the command would be ``\texttt{tcpdump -i ens160 -c 100}'' (see the inputs listed in Table~\ref{tab:largeApps}, Section~\ref{sec:evaluation}).
    The first argument identifies the Ethernet interface $\texttt{ens160}$, while the second argument specifies that $100$ packets should be captured.
    The former argument is a string (treated as an array of characters in C); the latter is an \texttt{int}.
    \Omit{\twr{Omit the following sentence: I don't think it adds anything:}}
    
    Returning to the example from \S\ref{sec:motivation} (Listing~\ref{wc}), Figure~\ref{fig:neck} illustrates the location of the neck.
    Suppose that the desired functionality is to count the number of lines (i.e., \texttt{wc -l}).
    Table~\ref{tab:vars} shows a subset of the variables and their corresponding values that will be captured and stored in \lmcas's database after partial interpretation finishes.
    \begin{table}[!htb]
    \centering
    \caption{Partial program state that contains the set of captured variables $V_s$ and the corresponding values obtained at the neck after partial interpretation of the program in Listing~\ref{wc}.}
    \scalebox{0.9}{
    \begin{tabular}{lllc}
        \hline
        Variable & Type & Scope & Value \\ \hline
        \texttt{total_lines} & \multirow{2}{*}{\texttt{int}} & \multirow{2}{*}{Global} & 0 \\
        \texttt{total_chars} &                   &                   & 0 \\ \hline
        \texttt{flag->count_lines} & \texttt{int} & \multirow{2}{*}{Local} & 1 \\
        \texttt{flag->count_chars} & \texttt{char} &                   & 0 \\ \hline
    \end{tabular}
    }
    \label{tab:vars}
    \end{table}

    
    
\subsection{Constant Conversion (CC)} \label{subsec:cc}

This phase aims to propagate constants in the configuration logic to enable further optimizations.
For instance, this phase contributes to removing input arguments that were not enabled during symbolic execution, and thus allows tests that check those inputs to be eliminated.

Constant conversion is a non-standard optimizing transformation because optimization is
performed \emph{upstream} of the neck:
uses of variables in the configuration logic---which comes \emph{before} the neck---are converted to constants, based on values captured at the neck.
The transformations performed during this phase enforce that the state at the neck in the debloated program is consistent with the partial state of constants at the neck that was captured at the end of partial interpretation.
Standard dataflow analyses (e.g. Def-Use)~\cite{Storm} for global and stack variables is used to replace all occurrences of the variables with their corresponding constant values in the program code before the neck.
Because some of the program statements become dead after constant conversion, the replacement is performed for all occurrences (i.e., accesses) of the variables obtained after partial interpretation.

The CC phase receives as input the bitcode of the whole program $P$ generated using WLLVM,\footnote{https://github.com/SRI-CSL/whole-program-llvm} as well as a dictionary (similar to Table~\ref{tab:vars}) that maps the set of variables in $V_s$ captured after symbolic execution to their constant values $C_s$. The set $V_s$ involves global and stack variables (base-type, struct, and pointer variables).
The CC phase then iterates over the IR instructions to identify the locations where the variables are accessed, which is indicated by load instructions.
Then, it replaces the loaded value with the corresponding constant value.
This approach works for global variables and stack variables with base types.
However, for pointers to base variables, it is necessary to identify locations where the pointer is modifying a base variable (by looking for store instructions whose destination-operand type is a pointer to a base type).
The source operand of the store operation is modified to use the constant value corresponding to the actual base variable pointed to by the pointer.

\rev{
For stack variables that are Structs and pointers to Structs, we first need to identify the memory address that is pointed to by these variables, which facilitates tracing back to finding the corresponding struct and pointer-to-struct variables.
We then iterate over the use of the identified memory addresses to determine store operations that modify the variable (corresponding to the memory addresses).
}
Finally, we convert the source operand of the store operations to the appropriate constant.
We also use the element index recorded during symbolic execution to identify which struct element should be converted.

For string variables, we identify the instructions that represent string variables, create an array, and assign the string to the created array. Finally, we identify store instructions that use the string variable as its destination operand, and override the store instruction's source operand to use the constant string value.

In \texttt{wc} (Listing~\ref{wc}), no replacements are performed for global variables \texttt{total\_lines} and \texttt{total\_chars} before the neck: there are no such occurrences.
Replacements are performed for referents of the pointer-to-struct \texttt{flag}:
the occurrences of $\texttt{flag->count\_chars}$ and $\texttt{flag->count\_lines}$ at lines 13 and 14 are replaced with the corresponding values listed in Table~\ref{tab:vars}.

\subsection{Multi-Stage Simplification (MS)} \label{subsec:msp}

This phase begins with the result of constant conversion, and performs whole-program optimization to simplify and remove unnecessary code.
In this phase, we used existing LLVM passes, as well as one pass that we wrote ourselves.
In particular, \lmcas uses the standard LLVM pass for constant propagation to perform constant folding;
it then uses another standard LLVM pass to simplify the control flow.
Finally, it applies an LLVM pass we implemented to handle the removal of unnecessary code. 

    \noindent \textbf{Constant Propagation}.
    This optimization step folds variables that hold known values by invoking the standard LLVM constant-propagation pass. Constant folding allows instructions to be removed.
    \Omit{
    It also opens up the opportunity for additional simplifications, such as ``Simplifying CFG'' below.} 
    
    \noindent \textbf{Simplifying the CFG}.
    \lmcas benefits from the previous step of constant propagation to make further simplifications by invoking a standard LLVM pass, called \texttt{simplifycfg}.
    This pass determines whether the conditions of branch instructions are always true or always false:
    unreachable basic blocks are removed, and basic blocks with a single predecessor are merged.
    \Omit{
    The simplifications in this pass are achieved at the basic-block level. }
    \begin{algorithm}[!htb]
\scriptsize
\caption{\lmcas Clean up}
\label{alg:cp}
\DontPrintSemicolon
\KwIn{$P_{cc}$, visitedFunc}
\KwOut{$P$\textprime}
    \SetKwProg{Def}{def}{:}{}
    $P\textprime\leftarrow P_{cc}$ \;
       \textcolor{blue}{\tcc{Remove unused functions}}
       CG $\leftarrow$ constructCallGraph($P'$) \;
       \For{$\textit{func} \in CG$}{\label{alg:l:funcloop1-b}
        \If{$\textit{func}\notin visitedFunc~\land$ $\textit{func}$ is not an operand of other instructions}{
            remove $\textit{func}$ from $P'$ and CG\;\label{alg:l:funcloop1-e}
        }
      }
      \For{$\textit{func} \in CG$}{\label{alg:l:funcloop2-b}
        \If{$\textit{func}$ is not an operand of other instructions}{
            remove $\textit{func}$ from $P'$ and CG \;
            remove $\textit{func}$'s descendent nodes from $P'$ and CG if they are not reachable from $\textit{main}$ \;\label{alg:l:funcloop2-e}
        }
      }
      \textcolor{blue}{\tcc{Remove unused Global Variables}}
      \For{$var \in getGlobalList(P_{cc})$}{\label{alg:l:gbls-b}
        \If{$var$ is not an operand of other instructions}{
            remove var from $P'$ \;\label{alg:l:gbls-e}
        }
      }
      \textcolor{blue}{\tcc{Remove unused Stack Variables}}
      \For{$\textit{func} \in CG$}{\label{alg:l:stack-b}
        \For{$inst \in \textit{func}$}{
            \If{$inst$ is $AllocInst$}{
                \If{$inst$ is not an operand of other instructions}{ 
                    remove inst from $P'$ \;
                }\uElseIf{$inst$ is a destination operand of only one $storeInst$}{ 
                    remove $storeInst$ from $P'$ \;
                    remove $inst$ from $P'$ \;\label{alg:l:stack-e}
                }
            }
        }
      }
\end{algorithm}

    \noindent \textbf{Clean Up}.
    In the simplification pass, \lmcas removes useless code (i.e., no operation uses the result~\cite{cooper2011engineering}) and unreachable code, including dead stack and global variables and uncalled functions.
    Although LLVM provides passes to perform aggressive optimization, we wrote a targeted LLVM pass that gives us more control in prioritizing the cleaning up of unneeded code, as described in Algorithm~\ref{alg:cp}, which receives the modified program $P_{cc}$ after the CC phase and the list of functions visited during the Partial Interpretation phase (\texttt{visitedFunc}). 
    
    The first priority is to remove unused functions.
    The goal is to remove two categories of functions:
    (i) those that are called only from call-sites before the neck, but not called during symbolic execution (Lines \ref{alg:l:funcloop1-b}-\ref{alg:l:funcloop1-e}), and (ii) those that are never called from the set of functions transitively reachable from \texttt{main}, including indirect call-sites (Lines \ref{alg:l:funcloop2-b}-\ref{alg:l:funcloop2-e}).
    Function removal is performed after constructing the call graph at Line 3.\Omit{\ma{I added the following for discussing indirect call sites}}
    To handle indirect-call sites, Algorithm~\ref{alg:cp} also checks the number of uses of a function, at Lines $5$ and $8$, before removing the function.
    This check prevents the removal of a function invoked via a function pointer.
    
    The focus then shifts to simplifying the remaining functions.
    For removing global variables (Lines \ref{alg:l:gbls-b}-\ref{alg:l:gbls-e}), we iterate over the list of global variables 
    and remove unused variables.
    Finally, we remove stack variables (Lines \ref{alg:l:stack-b}-\ref{alg:l:stack-e}), including initialized but unused variables by iterating over the remaining functions and erasing unused allocation instructions.
    (In general, standard LLVM simplifications do not remove a stack variable that is initialized but not otherwise used because the function contains a store operation that uses the variable.
    Our clean-up pass removes an initialized-but-unused variable by deleting the store instruction, and then the allocation instruction.)

In \texttt{wc} (Listing~\ref{wc}),
after the CC phase both the \texttt{count\_chars} and \texttt{count\_lines} fields of the struct pointed to by stack variable \texttt{flag} are replaced by the constants $0$ and $1$, respectively (see Table~\ref{tab:vars}).
The simplification steps remove the tests at lines $18$ and $20$ because the values of the conditions are always true.
Because the values of the conditions in the tests at lines $17$ and $19$ are always false, control-flow simplification removes both the tests and the basic blocks in the true-branches.
Furthermore, the removal of these basic blocks removes all uses of the global variable \texttt{total\_chars}, and thus the cleanup step removes it as an unused variable. 


%% file: neck.tex
\begin{figure}[tb]
    \centering
    \includegraphics[width=0.7\columnwidth]{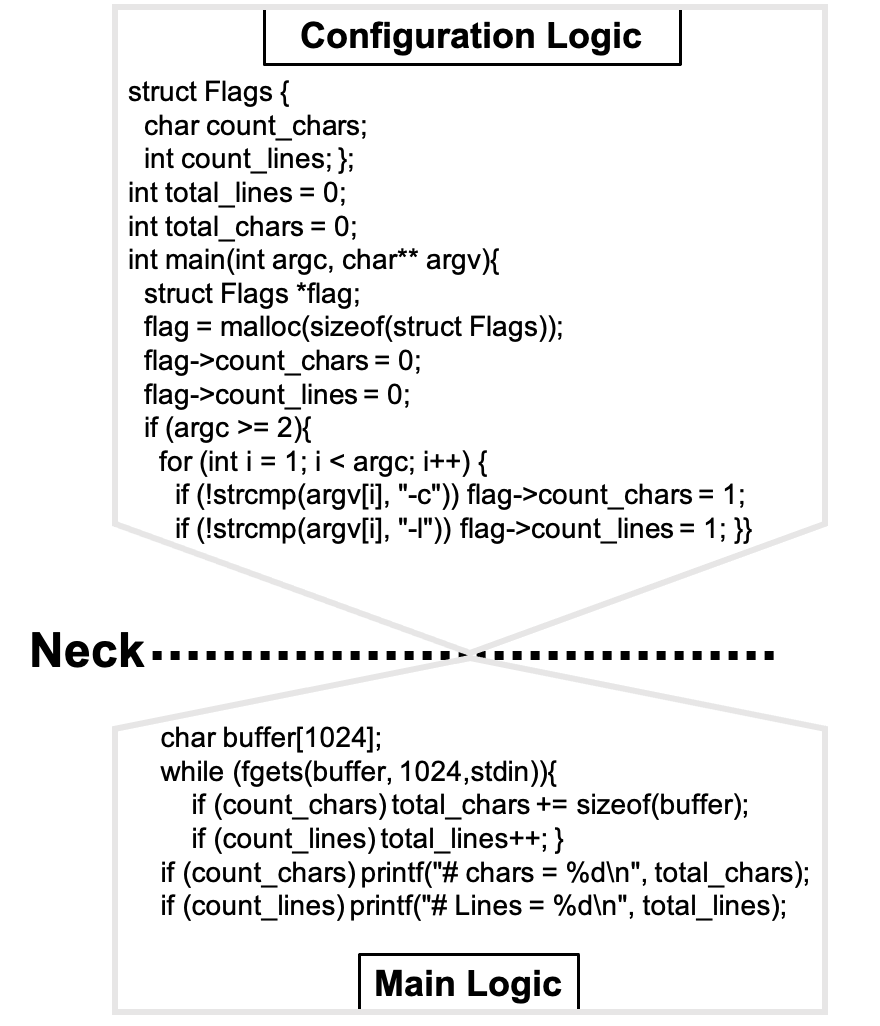}
    \caption{
    The neck miner selects Line $15$ as a splitting point to partition the motivating example in Listing~\ref{wc} into \textit{configuration logic} and \textit{main logic}. The splitting point is called the neck.
    }
    \label{fig:neck}
\end{figure}

\revUN{We developed a neck miner to recommend potential neck locations. To illustrate the neck idea, consider the example in Figure~\ref{fig:neck}, which represents the motivating example of Listing~\ref{wc}, but with the split between configuration logic and main logic emphasized.
The configuration-logic part consists of Lines 4-14, which include the declaration and initialization of global variables \texttt{total\_lines} and \texttt{total\_chars}, and the first part of function \texttt{main}.
The rest of \texttt{main} (Lines 15-20) represents the main logic, which contains the functionalities of counting lines and counting characters.
We call the point at the boundary between the two components the \emph{neck}.
Because the neck is located at the end of the configuration logic---where core arguments are parsed---the neck location is independent of the values of supplied inputs.
Thus, neck identification only needs to be conducted a single time for each program, and the neck location can be reused for different inputs---i.e., for different invocations of the debloater on that program. \revUN{According the the motivating example (Listing~\ref{wc}), the same neck location at Line $15$ can be used for debloating whether the supplied input is ``\texttt{wc -l}''  or ``\texttt{wc -c}'' .} 
}

\revUN{The neck miner uses two analyses: \emph{heuristic analysis} and \emph{structural analysis}, as described in Algorithm~\ref{alg:neckMiner}.
The \emph{heuristic analysis} relies on various patterns, corresponding to \emph{command-line} and \emph{configuration-file} programs, to identify a location from which to start the \emph{structural analysis}, which identifies the neck properly.
}

\begin{algorithm}[!htb]
\scriptsize
\caption{Neck Miner Algorithm}
\label{alg:neckMiner}
\DontPrintSemicolon
\KwIn{CFG, EntryPoint, programCategory, fileParsingAPIs}
\KwOut{NeckLocation}
    $startingPointForStructuralAnalysis = NULL$\ \;
    $distanceToNeckLoc \leftarrow \emptyset$\ \;
    $distanceToInst \leftarrow \emptyset$\ \;
      \textcolor{blue}{\tcc{Heuristic Analysis}}
        \If{programCategory is Command-Line}{ \label{alg:cmdLineBegin}
            \For{$inst \in \textit{UsesOf(argv)}$}{\label{alg:nm-trackargv}
                \If{$inst$ is inside a loop-structure}{ \label{alg:nm-insideLoop}
                    distance = computeDistance(inst,EntryPoint,CFG)\;
                    $distanceToInst \leftarrow distanceToInst \cup pair(inst,distance)$ \; \label{alg:cmdLineEnd}
                }
            }
        }\uElseIf{programCategory is Config-File}{ \label{alg:ConfigFileBegin}
            \For{$inst \in$ \textit{CFG}}{
                \If{$inst \in fileParsingAPIs$}{
                    distance = computeDistance(inst,EntryPoint,CFG)\;
                    $distanceToInst \leftarrow distanceToInst \cup pair(inst,distance)$ \; \label{alg:ConfigFileEnd}
                }
            }
        }
        $startingPointForStructuralAnalysis$ = InstAtShortestDistance(distanceToInst)\; \label{alg:nm-closest}
      \textcolor{blue}{\tcc{Structural Analysis}}
      \For{$inst \in CFG$}{\label{alg:nm-strtStructAnalysis}
        \If{inst is after startingPointForStructuralAnalysis in CFG}{\label{alg:nm-strtStructAnalysisFromScope}
            \If{inst satisfies the control-flow properties from \S\ref{Se:StructuralAnalysis}}{ \label{alg:nm-matchStmtProp}
                distance = computeDistance(inst,EntryPoint,CFG)\; \label{alg:nm-compDistNeck}
                $distanceToNeckLoc \leftarrow distanceToNeckLoc \cup pair(inst,distance)$ \; 
            }
        }
      }
      NeckLocation = InstAtShortestDistance(distanceToNeckLoc)\; \label{alg:nm-closestStmt}
      Add special function call before $NeckLocation$ to mark the neck\;
\end{algorithm}

\subsubsection{Heuristic Analysis}
\revUN{This step 
guides the structural analysis. It identifies a single location from which the structural analysis can be conducted, and relies on a set of patterns that apply to two categories of programs. 
These patterns are described as follows:}

\revUN{\textbf{Command-Line-Program Patterns (Algorithm~\ref{alg:neckMiner} Lines \ref{alg:cmdLineBegin}--\ref{alg:cmdLineEnd}):}
the inputs are provided to this category of programs via command-line arguments.
In C/C++ programs, command-line arguments are passed to \texttt{main()} via the parameters \texttt{argc} and \texttt{argv}:
\texttt{argc} holds the number of command-line arguments, and \texttt{argv[]} is a pointer array whose elements point to the different arguments passed to the program.
Consequently, this analysis step tracks the use of the argument \texttt{argv} (Line~\ref{alg:nm-trackargv}).
Specifically, because \texttt{argv} is a pointer array that, in general, points to multiple arguments, the analysis identifies the uses of \texttt{argv} that are inside of a loop (Line~\ref{alg:nm-insideLoop}). 
}

\revUN{\textbf{Configuration-File-Program Patterns (Algorithm~\ref{alg:neckMiner} Lines \ref{alg:ConfigFileBegin}--\ref{alg:ConfigFileEnd}):} this category of programs relies on configuration files to identify the required functionalities.
As a representative example, consider how the neck is identified in Nginx, a web-server program that supports $724$ different configuration options~\cite{EuroSec19}.
Listing~\ref{lst:nginx} presents a simple Nginx configuration file.
The gzip directive at line $8$ is associated with the libz.so library.
In some cases, multiple directives have to be defined to enable a certain capability, such as the SSL-related directives in lines $7$, $9$, and $10$ of Listing~\ref{lst:nginx}.
The heuristic analysis specifies the first location where the configuration file is parsed by certain APIs.
Identifying such APIs is simple because programs use system-call APIs to read files. 
For instance, nginx uses the Linux system call \texttt{pread}\footnote{https://man7.org/linux/man-pages/man2/pwrite.2.html} to read the configuration file.} 

Finally, after identifying the set of statements that match the various patterns, the heuristic analysis returns the statement that is closest to the CFG's entry point (Line~\ref{alg:nm-closest}), and ties are broken arbitrarily. In the motivating example (Listing~\ref{wc}), the heuristic analysis obtains the statement at Line~$13$ because it is the closest location to the entry point that matches the command-line patterns. 

\begin{lstlisting}[numbers=left,tabsize=2,
basicstyle=\footnotesize,keywordstyle=\color{blue},
numberstyle=\tiny,breaklines=true,numbersep=3pt, frame=tb,
caption=Nginx configuration file,captionpos=b
,keepspaces=true,xleftmargin=.03\textwidth, keywordstyle=\color{blue}\bfseries,
morekeywords={gzip,keepalive_timeout,server, access\_log, error\_log,worker_processes,charset,default_type,include,events,http,listen,ssl\_certificate,ssl\_certificate\_key},label=lst:nginx
]
worker_processes 1; 
events { worker_connections 1024; } 
http { 
	charset UTF_8;
	keepalive_timeout 65;
	server {
	 listen 443 ssl;               # libssl.so
	 gzip on; 				             # libz.so
	 ssl_certificate cert.pem; 	   # libssl.so
	 ssl_certificate_key cert.key; # libssl.so
	} }
\end{lstlisting}

\subsubsection{Structural Analysis}
\label{Se:StructuralAnalysis}
\revUN{This step identifies the neck location by analyzing the program's statements, starting from the location specified by the heuristic analysis (Lines~\ref{alg:nm-strtStructAnalysis}-~\ref{alg:nm-strtStructAnalysisFromScope} in Algorithm~\ref{alg:neckMiner}).
It identifies the statements that satisfy a certain set of control-flow properties that are discussed below (Line~\ref{alg:nm-matchStmtProp}).
Because it is possible to have several matching statements, the closest statement to the entry point is selected (Line~\ref{alg:nm-closestStmt}).
(Ties are broken arbitrarily.)
The closest statement is determined by computing the shortest distances in the CFG from the entry point to the neck candidates (Line~\ref{alg:nm-compDistNeck}).
The remainder of this section formalizes the aforementioned control-flow properties.
}

A program $P$ is a 4-tuple $(V,\textit{stmts},\textit{entry},\textit{exit})$, where $V$ is the set
of variables, $\textit{stmts}$ is the set of statements, $\textit{entry} \in \textit{stmts}$ is
the entry point of the program, and $\textit{exit} \in \textit{stmts}$ is the exit of
the program. 
As defined in Section~\ref{subsec:challenges}, we assume that there is a set $V_s \subseteq V$, which we call the set of {\it influenced variables} (e.g., command-line parameters of a utility).
Note that $V - V_s$ is the set of ``internal'' or non-influenced variables.
The location of a statement $s \in \textit{stmt}$ is denoted by $\textit{loc}(s)$. For simplicity, we assume that $\textit{Val}$ is the set of values that vars in $V$ can take.

Let $A: V_s \rightarrow Val \cup \{ \star \}$ be a partial assignment to the set of influenced variables (we assume that if $A(v) = \star$, then it has not been assigned a value). An assignment $A': V_s \rightarrow Val$ is {\it consistent} with partial assignment $A$ iff for all $v \in V_s$, if $A(v) \not= \star$, then $A'(v) = A(v)$. A statement $s \in \textit{stmt}$ is a {\it neck} for a program $P=(V,\textit{stmts},\textit{entry},\textit{exit})$ and a partial assignment $A$ (denoted by $\textit{nk}(P,A)$) if the following conditions hold:
\begin{itemize}
\item Given any assignment $A'$ consistent with $A$, $P$ always reaches the neck $\textit{nk}(P,A)$, and the statement corresponding to \textbf{the neck is executed exactly
once}.
This condition rules out the following possibilities:
Given $A'$, the execution of $P$ (i) might never reach the neck (the intuition here is that we do not want to miss some program statements, which would cause debloating to be unsound), or
(ii) the statement corresponding to the neck is inside a loop.


\item 
Let $R(nk(P,A))$ be all statements defined as follows: $s \in R(nk(P,A))$ iff $s$ appears after $nk(P,A)$, then $nk(P,A)$ could be \textbf{identified as articulation point} of the CFG of $P$ and one of the connected components
would over-approximate $R(nk(P,A))$. Another structural condition could be defined as follows: $R(nk(P,A))$ is \textbf{the set of all statements that are dominated by the neck}.

\Omit{
  \twr{No, "post-dominated" is not correct.
  I think you meant "dominated by the neck": we are talking about code downstream from the neck, so the neck is reached first.
  Separate from the problem definition: does the implemented algorithm impose the neck-domination condition?}\Omit{\ma{ Yes, as shown in the dom tree of the motivating example that I shared previously}}
}
  \Omit{\ma{should I put the dom tree that I generated  for our motivating example to prove that.}
  \twr{Possibly, but I think Figure 1 is sufficient.  I would put my point above (about sometimes needing a cut-set rather than a single cut-point) in Section 5.}}
  
\end{itemize}

\Omit{\twr{I don't understand the next sentence. The reader will wonder why the formalization allows there to be multiple possibilities.  Also, the what does it mean "the developer can determine [select?] the neck location" when you go on to say in the very next sentence that the neck location should be the closest to the entry point?} \ma{I rephrase the paragraph to clarify the exact manual step in the neck miner. To make it simple, all steps in the neck miner algorithm are automated except the step at line 19.}}
\revUN{The neck miner is fully automated, except the step at Line~\ref{alg:nm-matchStmtProp} (in Algorithim~\ref{alg:neckMiner}), which currently requires manual intervention.
We argue that such effort is manageable;
moreover, it is a one-time effort for each program.
Once the developer identifies the statements that satisfy the control-flow properties, they are fed to the neck miner to identify the statement that is closest to the entry point in the CFG.
Finally, a special function call (which serves to label the neck location) is inserted before the identified neck location. 
}

\Omit{\twr{This paragraph is inconsistent: you talk about "the developer iterating over the code", and in the next sentence "the miner selects".
The reader will wonder "Is the miner the developer or an algorithm?"}\ma{I rephrased the paragraph}}
\revUN{Consider Listing~\ref{wc} again.
The developer iterates over the program code, starting from Line $13$ (specified by the heuristic analysis) to identify the locations that satisfy the control-flow properties.
The developer ignores Lines $13$ and $14$ because they violate the control-flow properties:
they are not articulation points and are inside a loop, so not executed only once.
Line $15$ satisfies the control-flow properties because the statement at this location is executed only once, is an articulation point, and dominates all subsequent statements.
}

%% file: implementation.tex
\section{Implementation}

\noindent\textit{\textbf{Neck Miner}}.
\revUN{This component is implemented as an LLVM analysis pass.
In command-line programs, we use LLVM's def/use API to track the use of argv.
For configuration-file programs, we iterate over the LLVM IR code to identify call-sites for the pre-identified file-parsing APIs.
The developer has the responsibility of identifying the program locations that satisfy the structural properties from \S\ref{Se:StructuralAnalysis}.
(i.e., locations that are executed only once, and dominate the main logic).
This task is relatively easy because the developer can rely on existing LLVM analysis passes to compute the dominance tree and verify the structural properties.
We argue that such efforts are manageable. More importantly, they are one-time efforts.
(Such a semi-automated approach has also been used in prior work~\cite{mpi} and completely manual~\cite{Temporal}.)
Finally, the neck location is marked by adding a special function call to the program being analyzed.}

\noindent\textit{\textbf{Partial Interpretation}}.
\revUN{Our implementation uses KLEE~\cite{KLEE} to perform the partial interpretation because it (1) models memory with bit-level accuracy, and (2) can handle interactions with the outside environment---e.g., with data read from the file system or over the network---by providing models designed to explore various possible interactions with the outside world.}
We modified KLEE 2.1 to stop and capture the set of affected variables and their corresponding values after the neck is reached. 
\rev{In essence, KLEE is being used as an execution platform that supports “partial concrete states.”
For LMCAS, none of the inputs are symbolic, but only part of the full input is supplied.
In the case of word-count, the input is “wc -l”, with no file-name supplied. 
Only a single path to the neck is followed, at which point KLEE returns its state (which is a partial concrete state).
The second column in Tables \ref{tab:largeApps} and \ref{tab:benchmark1} describes the inputs supplied to KLEE for the different examples. 
}

\noindent\textit{\textbf{Multi-Stage Simplification.}}
\revUN{We also developed two LLVM passes using LLVM 6.0 to perform constant-conversion (CC) and the clean-up step of the MS phase.
We implemented these passes because of the absence of existing LLVM passes that perform such functionalities.
We tried existing LLVM passes like global dead-code elimination (DCE) to remove unused code.
However, global DCE is limited to handle only global variables (and even some global variables cannot be removed).
We also noticed that not all stack variables are removed, so in our clean-up pass we employ def-use information to identify stack variables that are loaded but not used.
Also, the removal of indirect calls is not provided by LLVM. 
To prevent the removal of functions invoked via a function pointer, our clean-up pass checks that the number of uses of a function is zero before removing the function.
}

%% file: evaluation_usenix.tex
\section{Evaluation}
\label{sec:evaluation}
This section presents our experimental evaluation of \lmcas. We address the following research questions:
\begin{itemize}
    \item \textbf{Effectiveness of Neck Miner:}
      How \textit{accurate} is the neck miner in identifying the neck location? (\ref{subsec:neckMiner})
    \item \textbf{Optimizing Debloating Process:}
      Does \lmcas \textit{speed up} the debloating process w.r.t. running time? (\ref{subsec:Optimizing})
    \item \textbf{Functionality Preserving and Robustness:} Does \lmcas produce \textit{functional} programs? and how \textit{robust} are the debloated programs produced by \lmcas? 
    (\ref{subsec:Preserving})
    \item \textbf{Code Reduction:}
      What is the debloating performance of \lmcas w.r.t. the amount that programs are \textit{reduced in size}? (\ref{subsec:Reduction})
    \item \textbf{Security:}
      Can \lmcas reduce the \textit{attack surface}? (\ref{subsec:security})
    \item \textbf{Scalability:}
      How \textit{scalable} is \lmcas in debloating large apps? (\ref{subsec:Scalability})
\end{itemize}

\noindent\textit{\textbf{Experimental Setup}}. Our evaluation relies on three datasets, as shown in Table~\ref{tab:dataset}.
\textit{Benchmark\_1} contains $15$ programs from GNU Coreutils v8.32 (see Table~\ref{tab:benchmark1}).
\textit{Benchmark\_2} contains six programs obtained from ChiselBench (see Table~\ref{table:cve}).\footnote{https://github.com/aspire-project/chisel-bench}
\textit{Benchmark\_3} consists of three programs (see Table~\ref{tab:largeApps}). 
\rev{
The selection of programs in Benchmark_1 was motivated by their use in prior papers on software debloating. We used Benchmark_2 because it provides us a list of CVE vulnerabilities and corresponding apps; considering this dataset facilitates our evaluation of the removal of CVEs, and allows us to compare against the CVE-removal performance of Chisel and RAZOR. 
}

All experiments were conducted on an Ubuntu $18.04$ machine with a 2.3GHz Quad-Core Intel i7 processor and 16GB RAM, except the fuzzing experiment, for which we used an Ubuntu $18.04$ machine with a 3.8GHz Intel(R) Core(TM) i7-9700T CPU and 32GB RAM.

\begin{table}[!tb]
\centering
\caption{Benchmark sets used in the evaluation.}
\scalebox{0.9}{
\begin{tabular}{llcl}
\hline
Source & Label & \# of apps \\ \hline 
\thead{GNU Coreutils 8.32} & Benchmark\_1 & 15 \\ \hline
\thead{CHISEL Benchmark} & Benchmark\_2 & 6 \\ \hline
\thead{Tcpdump \& GNU Binutils} & Benchmark\_3 & 3 \\ \hline
\end{tabular}
}
\label{tab:dataset}
\end{table}

\noindent\textit{\textbf{Compared tools and approaches}}. To evaluate the effectiveness of \lmcas, we compared with the following tools and approaches:
\begin{itemize}
    \item Baseline. We establish the baseline by compiling each app's LLVM bitcode at the -O2 level of optimization. This baseline approach was used in prior work~\cite{TRIMMER}.
    \item OCCAM~\cite{OCCAM}. The system most comparable to the approach used by \lmcas. However, OCCAM does not perform constant propagation, and thus omits a major component of partial evaluation. 
    \item CHISEL~\cite{Chisel}. It requires the user to identify wanted and unwanted functionalities and uses reinforcement learning to perform the reduction process. 
    \item RAZOR~\cite{RAZOR}. Similar to CHISEL, RAZOR relies on test cases to drive the debloating but incorporates heuristic analysis to improve soundness. RAZOR performs debloating for binary code, while the others operate on top of the LLVM IR code.
\end{itemize}
\revUN{We considered CHISEL and RAZOR as they represent state-of-the-art tools that are applying aggressive debloating techniques. While we selected OCCAM because it is a state-of-the-art partial evaluation tool and thus is the closest to \lmcas. Comparing with these various tools facilitates the verify the capabilities and effectiveness of \lmcas.}

\subsection{Effectiveness of the Neck Miner} \label{subsec:neckMiner}
\revUN{In this experiment, we measured the effectiveness of the neck miner in facilitating neck identification.
Our evaluation involved the $26$ programs specified in Table~\ref{tab:dataset}.
These programs belong to various projects: Coreutils, Binutils, Diffutils, Nginx and Tcpdump. For all $26$ programs, neck mining was successful, and the identified neck location was used to perform debloating.}

\revUN{For some programs, such as GNU \texttt{wc} and \texttt{date}, there were multiple candidate neck locations before the shortest-distance criterion was applied at Line~\ref{alg:nm-closestStmt} of Algorithm~\ref{alg:neckMiner}.
(Table~\ref{tab:neckMinerResults} in the Appendix~\ref{apndx:neckMinerEvaluation} contains the full set of results.)
}

\revUN{The neck location is inside the \texttt{main} function for the majority of the programs, except \texttt{readelf} and \texttt{Nginx}. 
With the help of the neck miner, it took only a few minutes for the one manual step (Line~\ref{alg:nm-matchStmtProp} of Algorithm~\ref{alg:neckMiner}) needed to identify the neck locations.
More specifically, for each program the analysis time for the heuristic analysis was $2$ seconds on average, and it took $5-10$ minutes to perform the manual part of structural analysis.
This amount of time is acceptable, given that neck identification is performed only once per program.
}

\revUN{As mentioned in Section~\ref{subsec:neckSpec}, the neck is identified only once for each program: the same neck can be used, regardless of what arguments are supplied.
To verify this aspect, we debloated various programs based on different supplied inputs.
For example, we debloated \texttt{sort} and \texttt{wc} based on $4$ and $5$ input settings, respectively, and (for each program) the same neck location was used with all debloating settings.
Similarly, a single neck location is used for multiple debloatings for each of the programs listed in Tables~\ref{tab:largeApps} and \ref{tab:capability} (Appendix~\ref{apndx:neckMinerEvaluation}).
}

\subsection{Optimizing Debloating Process} \label{subsec:Optimizing}
\begin{figure}[!htb]
    \centering
    \includegraphics[width=1\columnwidth]{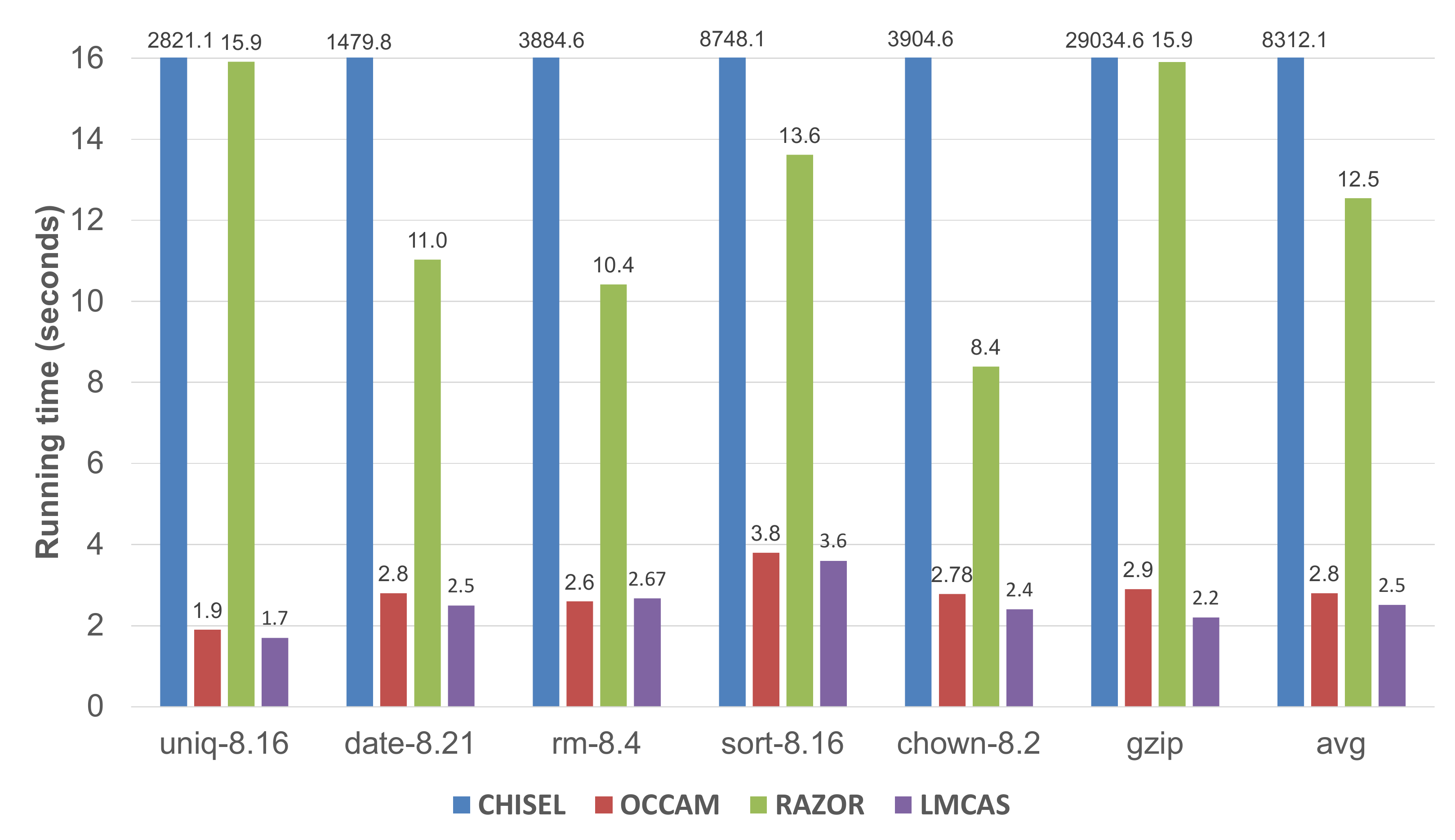}
    \caption{\Omit{\twr{The y-axis says "Analysis Time''.  Make it consistent with the text, which I changed to ``running time''.}} 
    Running times of \lmcas, CHISEL, RAZOR, and OCCAM based on \textit{Benchmark\_2} (ChiselBench).}
    \label{fig:analysisTimeComparison}
\end{figure}

\revUN{We compared the running time of \lmcas against those of CHISEL, RAZOR, and OCCAM on \textit{Benchmark\_2}. We used this benchmark because it was used by both CHISEL and RAZOR, thus the used test cases are available; otherwise, we need to come up with a set of test cases, which is not trivial. Debloating settings in this experiment are listed in Table~\ref{tab:chiselbenchinput}.
As depicted in Figure~\ref{fig:analysisTimeComparison}, the running times for \lmcas and OCCAM are significantly lower than the time for aggressive debloating techniques CHISEL and RAZOR. 
As a result, \lmcas runs up to $1500$x, $4.6$x, and $1.2$x faster on average than CHISEL, RAZOR, and OCCAM, respectively. This result illustrates \lmcas substantially speeds up the debloating process in contrast to aggressive debloating tools, but also slightly outperforms partial evaluation debloating techniques.}

\subsection{Functionality Preserving and Robustness} \label{subsec:Preserving}
\revUN{In this experiment, we ran the binaries before and after debloating against given test cases to understand their robustness. 
The majority of the programs debloated in \textit{Benchmark\_2} using RAZOR and CHISEL are suffering from run-time issues. The issues include crashing, infinite loop, or performing unexpected operations. These issues are reported and discussed in~\cite{RAZOR}. In our experiment, we found that all the debloated programs by CHISEL contain these issues. Among the programs in \textit{Benchmark\_2}, all but one of the OCCAM-debloated programs work correctly, and all of the \lmcas-debloated applications run correctly.}

\revUN{Since \lmcas and OCCAM have comparable results over \textit{Benchmark\_2}, we extended our evaluation by debloating programs in \textit{Benchmark\_1} according to the settings in Table~\ref{tab:benchmark1}. 
Five of the OCCAM-debloated programs ($33\%$) crash (i.e., segmentation fault) or generate inaccurate results, as reported in Table~\ref{tab:funcPreserving}. In contrast, all of the \lmcas-debloated programs run correctly (i.e., \lmcas preserves programs' behavior).}

We further assessed the robustness aspect of the debloated programs using fuzzing. It has been used to verify the robustness of debloated programs in~\cite{Chisel}. The aim was to test whether a debloated programs by \lmcas functioned correctly and did not crash. We used AFL (version 2.56b), a state-of-the-art fuzzing tool~\cite{afl}, to perform this experiment. We used AFL's black-box mode because our analysis is performed on LLVM bitcode; therefore, we could not use AFL to instrument the source code.
We ran AFL on the debloated programs created from \textit{Benchmark\_1} and \textit{Benchmark\_2} for six days. AFL did not bring out any failures or crashes of the debloated programs in either dataset. This experiment provides additional confidence in the correctness of the debloated programs created using \lmcas.

\begin{table}[]
\centering
\caption{Evaluation of functionality preserving after debloating by \lmcas and OCCAM for programs in Benchmark\_1. $\cmark$ means functionality is correctly preserved; otherwise: crashing (C), Infinite Loop(L), or Wrong operation (W).}
\scalebox{0.9}{
\begin{tabular}{|c|c|c|}
\hline
\multicolumn{1}{|l|}{\textbf{Program}} & \multicolumn{1}{l|}{\textbf{OCCAM}} & \multicolumn{1}{l|}{\textbf{LMCAS}} \\ \hline
basename                               & $\cmark$                               & $\cmark$  \\ \hline
basenc                                 & $\cmark$                               & $\cmark$  \\ \hline
comm                                   & $\cmark$                               & $\cmark$  \\ \hline
date                                   & $\cmark$                               & $\cmark$  \\ \hline
du                                     & W                                      & $\cmark$  \\ \hline
echo                                   & $\cmark$                               & $\cmark$  \\ \hline
fmt                                    & $\cmark$                               & $\cmark$  \\ \hline
fold                                   & $\cmark$                               & $\cmark$  \\ \hline
head                                   & L                                      & $\cmark$  \\ \hline
id                                     & $\cmark$                               & $\cmark$  \\ \hline
kill                                   & W                                      & $\cmark$  \\ \hline
realpath                               & $\cmark$                               & $\cmark$  \\ \hline
sort                                   & L                        & $\cmark$  \\ \hline
uniq                                   & $\cmark$                               & $\cmark$  \\ \hline
wc                                     & C                           & $\cmark$  \\ \hline
\end{tabular}
}
\label{tab:funcPreserving}
\end{table}

\subsection{Code Reduction} \label{subsec:Reduction}
\begin{figure}[!htb]
    \centering
    \includegraphics[width=1\columnwidth]{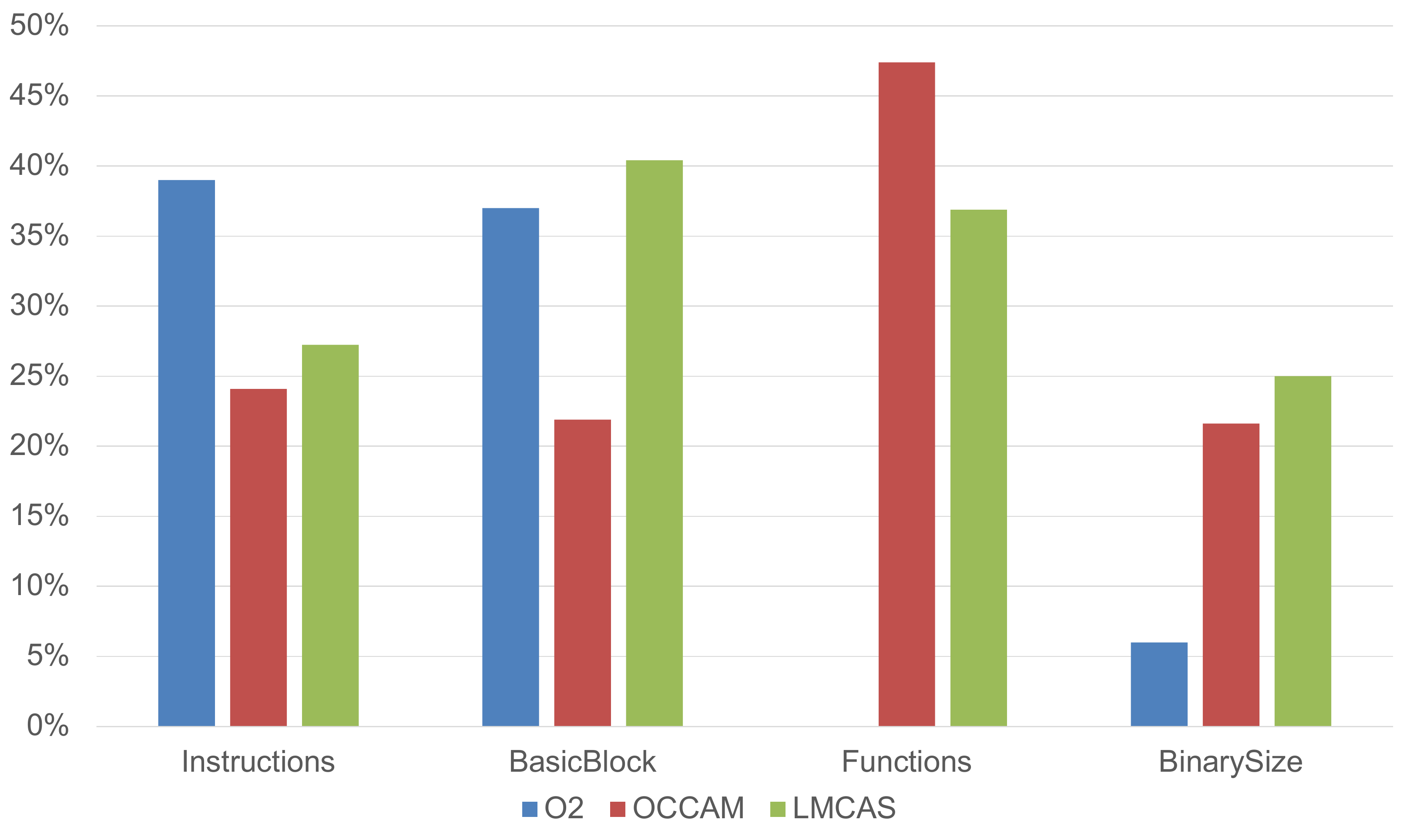}
    \caption{Average reduction in size achieved through baseline, OCCAM, and \lmcas, using four different size metrics. (Higher numbers are better.)}
    \label{fig:FineGrainedReductionO2_OCCAM_LMCAS}
\end{figure}

We used Benchmark\_1 to compare the performance of LMCAS against baseline and OCCAM. Figure~\ref{fig:FineGrainedReductionO2_OCCAM_LMCAS} shows the average reduction in size, using four different size metrics, achieved by baseline, OCCAM, and \lmcas.
All size measures, except binary size, are taken from the LLVM Intermediate Representation (IR). 
For computing the binary-size metric, we compiled all debloated apps with \texttt{gcc}, and ran \texttt{size}.
We report the sum of the sizes of all sections in the binary file (text + data + bss) because this quantity reflects the outcome of our simplifications across all sections. \lmcas achieved significant higher reduction rate (i.e., around the double) in comparison with baseline and OCCAM.

This result is due to the fact that the clean-up step of \lmcas can remove nodes in the call-graph that correspond to functions in binary libraries that are not used. 
Although the reduction rates of \lmcas and OCCAM are close (geometric mean binary size reduction is $25\%$ and $22\%$, respectively), some of the specialized programs generated by OCCAM are not reliable (as discussed in Section~\ref{subsec:Preserving}).

\begin{figure}[!htb]
    \centering
    \includegraphics[width=1\columnwidth]{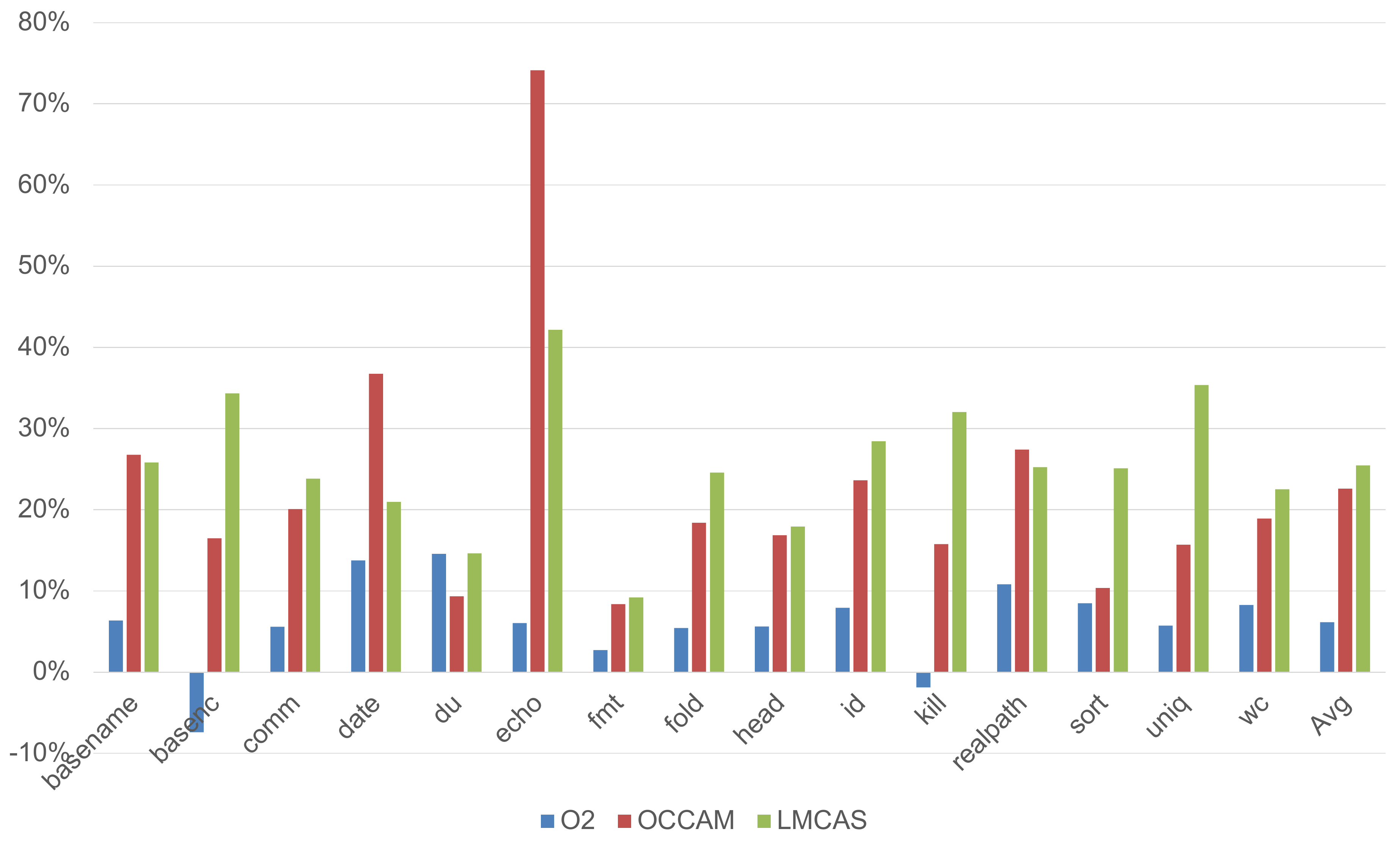}
    \caption{Binary size reduction achieved through baseline OCCAM, and \lmcas. (Higher numbers are better.)}
    \label{fig:binaryReductionO2_LMCAS_OCCAM}
\end{figure}

\revUN{Although the baseline shows a higher average reduction rate at instruction and basic block levels, its average reduction at binary size is the worse. Indeed, it increases the binary size for two programs (i.e., basenc and kill), as depicted in Figure~\ref{fig:binaryReductionO2_LMCAS_OCCAM} (Appendix~\ref{sec:additionalResultsComparison} presents extended results), which shows the comparison results based on the reduction in the binary size that each tool achieved for each app in \textit{Benchmark\_1}.} 

\subsection{Security benefits of \lmcas} \label{subsec:security}

We evaluated the capabilities of \lmcas to reduce code-reuse attacks and to remove known vulnerabilities. Therefore, we conducted the following three experiments: (i) we attempt to reproduce (by executing the app) the vulnerability after debloating the apps to see if the vulnerabilities were eliminated;
(ii) we measured the reduction in the number of code-reuse attacks\Omit{attack surface} by counting the number of\Omit{return-oriented programming (ROP)} eliminated gadgets in compiled versions of the original and reduced programs;
(iii) we compared  the degree of gadget reduction achieved by \lmcas and LLVM-CFI. 

\noindent\textbf{Vulnerability Removal}.
To test the ability to mitigate vulnerabilities, we used six programs in \textit{Benchmark\_2} because this benchmark contains a set of known CVEs.
Table~\ref{table:cve} presents a comparison between \lmcas, RAZOR, OCCAM, and CHISEL.
\lmcas and OCCAM removed CVEs from $5$ out of the $6$ programs, but the debloated \texttt{sort-8.16} using OCCAM was not behaving correctly at run-time. While CHISEL and RAZOR removed CVEs from $4$ programs. Although OCCAM removed CVEs from $3$ programs including \texttt{rm-8.4}, but the debloated \texttt{rm-8.4} shows unexpected infinite-loop behaviour. We suspect that OCCAM may remove loop condition checks.   
\lmcas could not remove the vulnerability in \texttt{date-8.21} because the bug is located in the core functionality of this program.
When undesired functionality is too intertwined with the core functionality---e.g., multiple quantities are computed by the same loop---then \lmcas may not be able to remove some undesired functionality because---to the analysis phases of \lmcas---it does not appear to be useless code.
In such cases, \lmcas retains the desired functionality. 
In contrast, CHISEL tries to remove all undesired functionality.
\begin{table}[!tb]
\centering
\caption{Vulnerabilities after debloating by \lmcas and CHISEL. $\cmark$ means the CVE vulnerability is eliminated; $\xmark$ means that it was not removed.}
\scalebox{0.75}{
\begin{threeparttable}
\begin{tabular}{llcccc}
\hline
App & CVE ID & RAZOR & CHISEL & OCCAM & \lmcas \\ \hline \hline
chown-8.2 & CVE-2017-18018 & $\cmark$ & $\cmark$ & $\cmark$ & $\cmark$ \\ \hline
date-8.21 & CVE-2014-9471 & $\cmark$ & $\xmark$ & $\xmark$ & $\xmark$ \\ \hline
gzip-1.2.4 & CVE-2015-1345 & $\cmark$ & $\cmark$ & $\xmark$ & $\cmark$ \\ \hline
rm-8.4 & CVE-2015-1865 & $\xmark$ & $\xmark$ & $\xmark$ & $\cmark$ \\ \hline
sort-8.16 & CVE-2013-0221 & $\cmark$ & $\cmark$ & $\cmark$ & $\cmark$ \\ \hline
uniq-8.16 & CVE-2013-0222 & $\xmark$ & $\cmark$ & $\cmark$ & $\cmark$\\ \hline
\end{tabular}
 \begin{tablenotes}
     \item[1] \small \revUN{CHISEL and RAZOR CVE removal is obtained from the corresponding publication.
     \item[2] \small Although OCCAM removed the CVE in rm-8.4, but the debloated version suffers from infinite loop at run-time.}
   \end{tablenotes}
  \end{threeparttable}
}
\label{table:cve}
\end{table}




\noindent\textbf{Gadget Elimination}.
Code-reuse attacks leverage existing code snippets in the executable, called gadgets, as the attack payload~\cite{CSET19}.
Those gadgets are often categorized---based on their last instruction---into Return-Oriented Programming (ROP), Jump-Oriented Programming (JOP), and syscall (SYS)~\cite{Piece-Wise,CARVE}.
Code-reuse attacks can be mitigated using software diversification, which can be achieved through application specialization.
Software debloating is a specialization technique that can both reduce the number of exploitable gadgets and change the locations of gadgets, thus diversifying the binary. 

\begin{figure}[htb]
    \centering
    \includegraphics[width=\columnwidth]{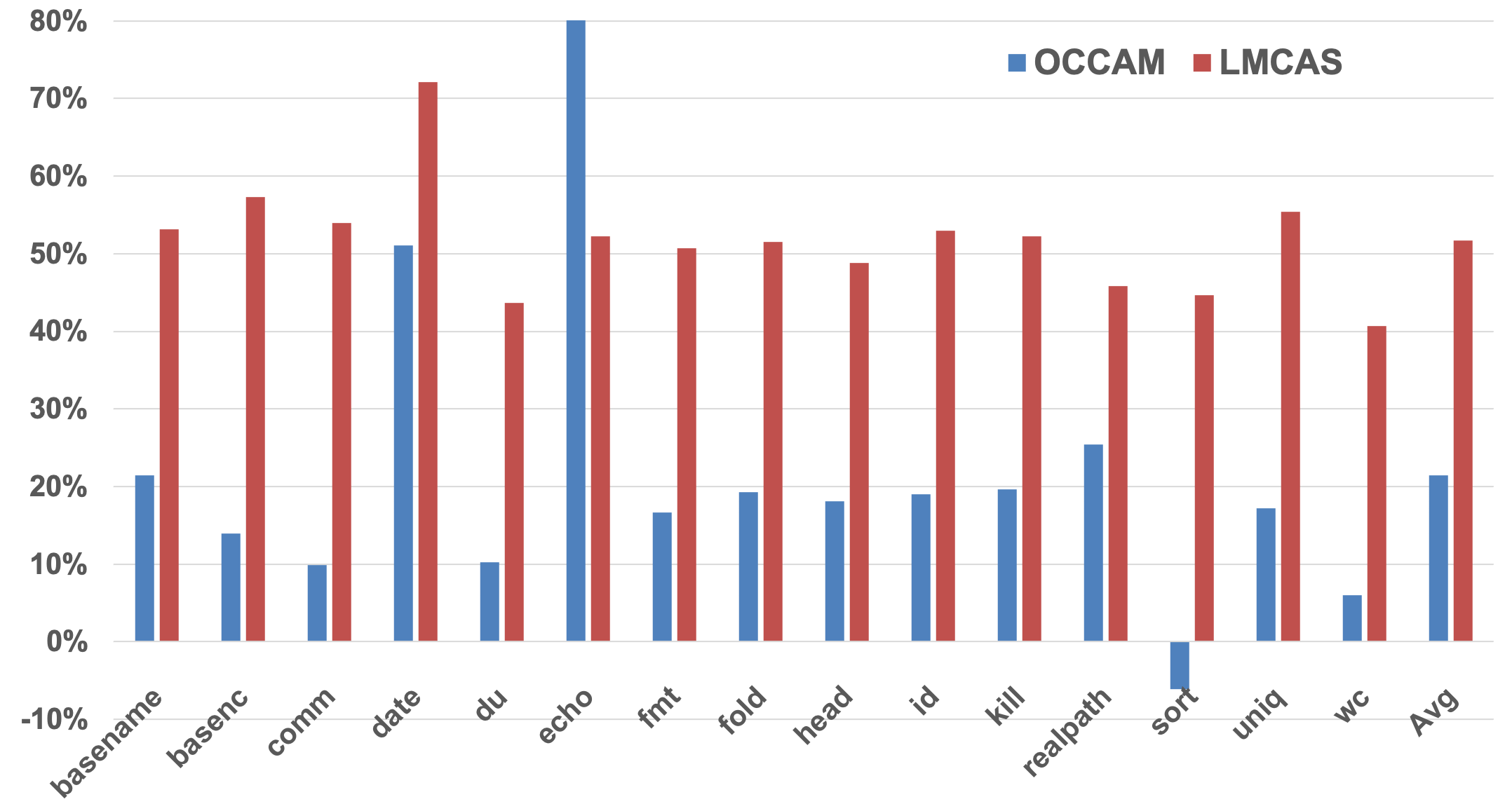}
    \caption{Reduction in the total number of unique gadget occurrences (SYS, ROP, and JOP) between \lmcas and OCCAM for \textit{Benchmark\_1}. (Higher numbers are better.)}
    \label{fig:rop}
\end{figure}

In this experiment, we used \textit{Benchmark\_1} and observed noticeable reductions in the total gadget count \Omit{\twr{What are SYS and JOP? Are you saying that you have two other figures like Figure~\ref{fig:rop} for two other kinds of gadgets?  If so, break that out into a separate paragraph (after this one), and report an aggregate number like the geometric means for the SYS and JOP measurements.  Have I explained the subtleties of computing geometric mean of percentages to you?  You cannot just compute the geometric means of the raw percentage numbers.} \ma{I apologies for the incorrect description, but Fig 6 reports the 3 types of gadgets SYS, JOP and ROP. I fixed that by making sure everything is consistent that we report total gadgets}}(occurrences of ROP, SYS, and JOP gadgets), as illustrated in Figure~\ref{fig:rop}.
We used ROPgadget~\cite{ROPGadget} for counting the number of gadgets.
\Omit{\twr{I think we need to clarify what is meant by the phrase ``total number of gadgets.''
From hidden comments, you say you mean SYS + JOP + ROP gadgets.
However, there is still ambiguity about whether you are counting
gadget \emph{kinds} or gadget \emph{instances}.
That is if the program contains 5 instances of gadget g1, 3 of gadget g2,
and 12 of gadget g3, is that ``three total gadgets'' or ``twenty total gadgets''?
} \ma{20 if all gadgets are unique}}
\Omit{\twr{How is average computed?  Say which kind of average you are reporting: arithmetic mean or geometric mean.} \ma{Arithmetic}}The average reduction (arithmetic mean) in the total number of unique gadgets is $51.7\%$ and the maximum reduction is $72.1\%$ (for \texttt{date}), while OCCAM reduces the total number of unique gadgets by $21.5\%$, on average, with a maximum reduction of $80.3\%$ (for \texttt{echo}).
For one program, \texttt{sort}, OCCAM increases the total number gadgets by $6.2\%$.
 With \lmcas, the number of SYS gadgets is reduced to $1$ for $14$ out of the $15$ applications.
\lmcas caused an increase in the number of SYS gadgets in one application (\texttt{sort}), but still produced an overall decrease when considering ROP and JOP. A similar increase is also observed with TRIMMER in three applications~\cite{TRIMMER}.

\noindent
\revUN{\textbf{Control-Flow Integrity (CFI).}
CFI is a prominent mechanism for reducing a program's attack surface by preventing control-flow hijacking attacks:
CFI confines the program to some specific set of CFG paths, and prevents the kinds of irregular transfers of control that take place when a program is attacked.
Although CFI does not specifically aim to reduce the number of gadgets, others have observed empirically that CFI reduces the number of unique gadgets in programs \cite{enforce_CFI,Fine-CFI,Ancile}.
Thus, we compared the degree of gadget reduction achieved by \lmcas and LLVM-CFI (a state-of-the-art static CFI mechanism)~\cite{Ancile}.
We compiled the LLVM bitcode of our suite of programs using \texttt{clang}, with the flags \texttt{-fsanitize=cfi -fvisibility=default}.
Among the $20$ programs analyzed, \lmcas outperformed LLVM-CFI on $12$ programs (60\%) by creating a program with a smaller total number of unique gadgets. (Table~\ref{tab:cfiAnalysis} in the Appendix~\ref{apndx:cfi} contains the full set of results.)
The last column in Table~\ref{tab:cfiAnalysis} shows that a significant reduction in unique gadgets---beyond what either \lmcas or LLFM-CFI is capable of alone---is obtained by first applying \lmcas and then LLVM-CFI.}

\subsection{Scalability} \label{subsec:Scalability}
We evaluated the capability of \lmcas to handle \textit{large-scale programs} that maintain complex input formats, such as object files, binaries, and network packets.

We used \textit{Benchmark\_3} in this experiment. The programs considered in \textit{Benchmark\_3} have been used in prior work to evaluate scalability~\cite{OCCAM,SymCC}, including the scalability of KLEE~\cite{moklee}.
Accordingly, we use the following applications\footnote{
  Because deciding which lines of code belong to an individual application is difficult, lines of code (LOC) are for the whole application suites from which our benchmarks are taken.
  We used \texttt{scc} (https://github.com/boyter/scc) to report LOC.
}
to show the scalability of \lmcas: 
\begin{itemize}
    \item \texttt{tcpdump}~\cite{tcpdump} (version 4.10.0; 77.5k LOC) analyzes network packets.
    We link against 
    its accompanying libpcap library (version 1.10.0; 44.6k LOC).
    \item \texttt{readelf} and \texttt{objdump} from GNU Binutils~\cite{Binutils} (version 2.33; 78.3k LOC and 964.4k of library code\footnote{
      We only consider lines in the binutils folder and the following dependencies: libbfd, libctf, libiberty, and libopcodes.
      }).
      \texttt{readelf} displays information about ELF files, while \texttt{objdump} displays information about object files.
\end{itemize}

We debloated these programs using different inputs to illustrate the capability of \lmcas to handle and debloat programs based on single and multiple inputs.
For example, we debloated \texttt{readelf} based on one argument \texttt{(-S)} and nine arguments \texttt{(-h -l -S -s -r -d -V -A -I)}. Table~\ref{tab:largeApps} breaks down the analysis time in terms of Partial Interpretation (third column) and the combination of Constant Conversion and Multistage Simplification (fourth column).
For these programs, the time for symbolic execution is lower than that for the LLVM-based simplifications.
This situation is expected because these programs contain a large number of functions, so a longer time is needed for the LLVM simplifications steps.
The inclusion of third-party libraries diminishes the reduction rate in the binary size, which is clearly illustrated in the achieved reduction rate for \texttt{tcpdump}. 

\begin{table}[!tb]
\centering
\caption{Scalability analysis of large applications, based on various inputs.
\Omit{\twr{This table is now inconsistent with Figure \ref{fig:analysisTimeComparison}: one refers to CC \& MS; the other refers to "partial-evaluation steps".  I prefer CC \& MS because it is more specific.} \ma{I fixed that, everything now is based on CC \& MS}} 
}
\small
\scalebox{0.9}{
\begin{tabular}{|p{1.2cm}|p{1.8cm}|p{0.8cm}|p{1.2cm}|p{1.9cm}|}
\hline
               Program & Supplied Inputs & PI (sec) & CC \& MS (sec)& Binary Size\newline Reduction Rate\\ \hline
\multirow{2}{*}{tcpdump} & -i ens160 & 48.1 & 173.1 & 2\%\\ \cline{2-5}
                  & -i ens160 -c 5 & 48.2 & 201.7 &2\% \\ \hline
\multirow{2}{*}{readelf} & -S & 10.6 & 41.7 & 4.8\%\\ \cline{2-5}
                  & -h -l -S -s -r -d -V -A -I & 20.15 & 72.4 & 4.71\%\\ \hline
                 \multirow{2}{*}{objdump} & -x & 40.84 & 246.17 & 5.65\%\\ \cline{2-5}
                 & -h -f -p & 48.07 & 320.11 & 5.71\%\\ \hline
                 
\end{tabular}
}
\label{tab:largeApps}
\end{table}

%% file: discussion_V2.tex
\section{Discussion and Limitations} \label{sec:discussion}

\rev{
\noindent\textbf{Generality of the neck concept.}
This concept applies to various types of programs. Our evaluation involves $26$ command-line programs, in which the neck can be easily identified, without discarding any program because the neck identification was not possible. We also inspected \texttt{Nginx} and observed that ALL directives in the configuration file are read before reaching the main logic (as in command-line programs). But this partitioning approach into configuration and main logic cannot be applied to Event-driven programs, which require constant interaction with the user to handle the functionalities required by the user.
Therefore, the configuration of program features is performed at various locations. However, we foresee our partitioning approach applies to Event-driven programs that apply server architecture, whose life-cycle is divided into initialization and serving phases~\cite{Temporal}.  
Our future work will consider evaluating such programs. 
}

\noindent\revUN{\textbf{Incorrect neck identification}. 
Misidentifying the neck may lead to incorrect debloating. But the neck miner incorporates a set of heuristics and structural features to recommend accurate neck locations. The heuristic analysis techniques aid the neck identification by pinpointing the starting point where the neck miner establishes its analysis. For instance, the GNU Coreutils programs use a particular idiom for parsing command-line parameters; in principle, a special-purpose algorithm could be designed for identifying the neck for programs that use that idiom. Then the neck miner applies a set of structural requirements to constrain the neck candidates. For example, the properties that the neck is executed only once and the neck is an articulation point. These structural properties reflect the nature of the neck definition.}
    
\noindent\textbf{Precision of Constant Conversion}.
    \lmcas relies on converting a subset of the variables in the captured concrete state into constants.
    If one of the variables could have been converted to a constant, but was not identified by \lmcas as being convertible (and therefore no occurrences of the variable are changed), no harm is done: that situation just means that some simplification opportunities may be missed.
    On the other hand, if some variable occurrences are converted to constants unsoundly, the debloated program may not work correctly.
    
    We mitigate this issue by: (1) avoiding the conversion of some variables to constants (e.g., \texttt{argv}) because it carries out the rest of inputs (i.e., delayed inputs) different than the one required by the specialized program: for instance, with \texttt{wc} the file name is not supplied during partial interpretation, while the file name is supplied to the debloated program; 
    (2) leveraging existing LLVM APIs (i.e., \texttt{getUser}) that track the uses of variables to capture the final constant values at the end of the partial interpretation.
    This approach overcomes situations where a pointer indirectly updates the value of a location, and ensures that the pre-neck constant-conversion step operates using updated and accurate constant values.

\noindent\revUN{\textbf{Reducing the Attack Surface}.
\lmcas reduces the attack surface in various ways by removing some known CVEs and eliminating some code-reuse attacks.
\Omit{\twr{The next sentence doesn't mean anything: remove. At best, one can say that the capabilities of \lmcas and LLVM-CRI are incomparable.} \ma{agreed}}
Our future work will enforce stronger security properties: specializing control-flow by restricting the set of allowed indirect control-flow transfers in the remained code by incorporating CFI techniques~\cite{Ancile,enforce_CFI} \Omit{\twr{What does it mean to tighten CFI checks (beyond the constraints that CFI already enforces)?}} and disabling security-critical system calls that are unused in the main logic~\cite{Temporal}.\Omit{\twr{What does it mean to "remove ... calls that are unused"? Instead of "remove" do you mean "disable"?  Or will the disabling be done inside the OS kernel, in which case what is \lmcas contributing (maybe just information about which system calls to disable)?} \ma{thanks for the correction, I meant disabling. Yes, this would be the contribution, but we already do this when we check the use of functions before removing them, therefore, what we need to do know, just capturing precisely the system calls.}}}


%% file: relatedWork.tex
\section{Related Work} \label{sec:rw}

A variety of software-debloating techniques have been developed in the research community, mainly over the last three years~\cite{EuroSec19,pldi10,stochasticOptimization,Temporal,RAZOR,BinaryTrimming,BinRec,CARVE,wedDebloat,RedDroid,Nibbler,ESORICS2019,Cozart,Unikernel}. In this section, we discuss various lines of research on software debloating and partial evaluation that are related to our work.

\noindent\revUN{\textbf{Program Partitioning.} 
Our work is peripherally related to prior work on program partitioning. 
MPI~\cite{mpi} reduces the overhead of provenience tracking and audit logging by portioning programs' execution based on annotated data structures. 
Privtrans~\cite{Privtrans} applies program partitioning to integrate privilege separation. Programs are divided into two components: the monitor and slave. These components run as separate processes, but cooperate to perform the same function as the original program. 
Glamdring~\cite{Glamdring} partitions applications at the source-code level, minimizing the amount of code placed inside an enclave. Our use of partitioning is for a different purpose: we create exactly two partitions, separated by the neck, and the partition before the neck is used to identify constant values that are then used to specialize the partition that comes after the neck.
\Omit{\twr{I'd like to hear Somesh's opinion of the above discussion.
To me it seems wrong to say that our motivation came from previous work on program partitioning.
That work primarily exploits multiple processes to introduce a degree of protection, which is not what we are doing at all.
One might be able to say that the \emph{analyses} that we do are inspired by program partitioning, but I'm not even sure that that is a correct characterization of our analyses.
Overall, I don't see the point of making such a connection, unless you think that it is easy to make a contrast between work in that area and our work.
But the current text is not bringing out points of contrast.}}
}

\revUN{In the debloating domain, Ghavamnia et al.~\cite{Temporal} propose a debloating approach to reduce the attack surface in server applications. This approach partitions the execution life-cycle of server programs into initialization and execution phases. It then reduces the number of system calls in the execution phase. However, this approach requires manual intervention from the developer to identify the boundary between the two phases, but without providing certain specifications to guide the identification process. In contrast, \lmcas performs specialization for the main logic and incorporates a neck miner to suggest a possible neck location. The neck miner provides semi-automatic support for the partitioning process and identified the neck correctly in $26$ programs.
}
\Omit{\twr{It seems imperative that we have a well-written comparision with Ghavamnia et al.
Does the work by Ghavamnia et al.\ do \emph{specialization} of the execution-phase code, or does it merely disable certain system calls.
If the latter, then the points of comparision are (i) \lmcas performs specialization, and (ii) \lmcas incorporates a neck miner to suggest a set of possible neck locations.
However, this comparison raises some questions that I don't think we address in Section \ref{sec:evaluation}:
(1) How well does the neck miner work---e.g., (a) for what percentage of the examples was the neck we used in the set returned by the neck miner?
(b) What is the average size of the set returned by the neck miner?
Also, ``avoid manual efforts'' is a weak phrasing;
it would be better to replace it with something like ``\lmcas provides semi-automatic support for the partitioning process by providing a neck miner that ... <some quick summary of success statistics for the neck miner>''
Finally, there is another issue that we don't have time to explore: does \lmcas have the same notion of ``initialization phase'' as Ghavamnia et al.?
That is, could our neck miner be applied to the examples used by Ghavamnia et al.\ to find the partitioning points that they used?
}}

\noindent\textbf{Partial Evaluation} has been used in numerous domains, including software debloating~\cite{TRIMMER}, software verification~\cite{Interleaving}, and to generate test cases~\cite{Albert2010PETAP}. Bubel et al.~\cite{Interleaving} use a combination of partial evaluation and symbolic execution. Because their capabilities are complementary, the two techniques have been employed together to improve the capabilities of a software-verification tool~\cite{Interleaving}. However, the goals and modes of interaction are different:
in the work of Bubel et al., partial evaluation is used to speed up repeated execution of code by a symbolic-execution engine; in our work, symbolic execution is in service to partial evaluation by finding values that hold at the neck. 

\noindent\textbf{Application Specialization}.
For debloating Java programs, JShrink~\cite{JShrink} applies static and dynamic analysis.  Storm is a general framework for reducing probabilistic programs~\cite{Storm}. For debloating C/C++ programs TRIMMER~\cite{TRIMMER} and OCCAM~\cite{OCCAM} use partial evaluation.
While TRIMMER overcomes the limitations of OCCAM by including loop unrolling and constant propagation. But in both tools, constant propagation is only performed for global variables, and thus TRIMMER and OCCAM miss specialization opportunities that involve local variables, which makes the debloating process unsafe. 
\lmcas can accurately convert the elements of struct variables into constants.
Furthermore, our analysis considers pointers, both to base types and struct types, which boosts the reliability of \lmcas. 


\revUN{Aggressive debloating tools like CHISEL~\cite{Chisel} and RAZOR~\cite{RAZOR} can achieve a significantly higher reduction rate in the size of specialized applications; however, these tools are prone to run-time issues (e.g., crashing, infinite loops). Furthermore, the debloating process takes a long time because these tools apply burdensome techniques, based on extensive program instrumentation and requiring users to provide a comprehensive set of test cases. RAZOR uses a best-effort heuristic approach to overcome the challenge of generating test cases to cover all code. While \lmcas applies lightweight techniques using program partitioning. Also, the specialized programs generated by \lmcas do not suffer from run-time issues.}

\noindent\textbf{Function Specialization}.
Saffire~\cite{Saffire} specializes call-sites of sensitive methods to handle certain parameters based on the calling context. 
\Omit{\twr{What does "Piecewise approach" mean?} \ma{it's the name of the tool. My wording was wrong}}Quach et al.~\cite{PieceWise} proposes a a tool called Piecewise for debloating libraries. Piecewise constructs accurate control-flow-graph information at compilation and linking time according to an applications’ usage.

%% file: conclusion.tex
\section{Conclusion}
\label{sec:Conclusion}

In this paper, we present \lmcas, a practical and lightweight debloating approach for generating specialized applications.
To speed up our analysis, \lmcas introduces the neck concept, a splitting point where ``configuration logic'' in a program hands off control to the ``main logic'' of the program. We develop a neck miner for alleviating the amount of manual effort required to identify the neck. \lmcas only applies partial interpretation up to the neck. The main logic is then optimized according to the values obtained from analyzing the configuration logic. Therefore, \lmcas eliminates the overhead of demanding techniques and boosts the safety of debloating process. \lmcas achieves a substantial reduction in the size of programs, and also reduces the attack surface.

%% file: appendix.tex
\appendix
\section{Benchmark Characteristics and Debloating Settings} \label{sec:chiselSettings}
Table~\ref{tab:benchmark1} lists the programs in Benchmark\_1 and the supplied inputs for debloating, and mentions various size metrics about the original programs. Table~\ref{tab:chiselbenchinput} provides the list of supplied inputs that we used for debloating the programs in Benchmark\_2. This list of inputs are obtained from~\cite{RAZOR}.
\begin{table}[!h]
\centering
\small
\caption{Characteristics of the benchmarks in \textit{Benchmark\_1}.}
\scalebox{0.9}{
\begin{tabular}{|p{1.3cm}|p{1.55cm}|p{0.8cm}|p{0.65cm}|p{1.1cm}|p{1cm}|}
\hline
\multirow{2}{*}{Program} & \multirow{2}{*}{\shortstack{Supplied\\ Inputs}} &  \multicolumn{4}{c|}{Original} \\\cline{3-6}
 &  & \shortstack{\# IR \\ Inst.} & \shortstack{\#\\ Func.}  & \shortstack{\# Basic\\ Blocks} & \shortstack{Binary \\ Size $^a$} \\ \hline
basename & \-\-suffix=.txt & 4,083 & 96 & 790 & 26,672 \\ \hline
basenc & base64 & 8,398 & 156 & 1,461 & 44,583 \\ \hline
comm & -12 & 5,403 & 110 & 972 & 32,714 \\ \hline
date & -R & 29,534 & 166 & 6,104 & 89,489 \\ \hline
du & -h & 5,0727 & 466 & 8,378 & 180,365 \\ \hline
echo & -E & 4,095 & 89 & 811 & 27,181 \\ \hline
fmt & -c & 5,732 & 115 & 1,095 & 79,676 \\ \hline
fold & -w30 & 4,623 & 100 & 893 & 29,669 \\ \hline
head & -n3 & 6,412 & 119 & 1,175 & 37,429 \\ \hline
id & -G & 5,939 & 125 & 1,172 & 36,985 \\ \hline
kill& -9 & 4,539 & 96 & 898 & 31,649 \\ \hline
realpath & -P & 8,092 & 155 & 1,419 & 41,946 \\ \hline
sort & -u & 25,574 & 329 & 3,821 & 116,119 \\ \hline
uniq & -d & 5,634 & 115 & 1,092 & 37,159 \\ \hline
wc & -l & 7,076 & 130 & 1,219 & 41,077 \\ \hline
\end{tabular}}
\label{tab:benchmark1}
\footnotesize{$^a$Total binary size obtained via the GNU \texttt{size} utility.}
\end{table}

\begin{table}[!h]
\centering
\caption{Input settings for the programs in \textit{Benchmark\_2} (obtained from~\cite{RAZOR}).}
\begin{tabular}{|c|c|}
\hline
Program & Supplied Inputs           \\ \hline
chown   & -h, -R                     \\ \hline
date    & -d, –rfc-3339, -utc        \\ \hline
gzip    & -c\\ \hline
rm      & -f, -r                     \\ \hline
sort    & -r, -s, -u, -z             \\ \hline
uniq    & -c, -d, -f, -i, -s, -u, -w \\ \hline
\end{tabular}
\label{tab:chiselbenchinput}
\end{table}

\section{Neck Miner Evaluation}\label{apndx:neckMinerEvaluation}
Table~\ref{tab:neckMinerResults} describes the neck miner evaluation results. The second column indicates whether multiple neck locations are matching the control-flow properties. The third column describes if the location of the selected neck location is inside the main function. 

Table~\ref{tab:capability} shows \lmcas performed debloating based on various debloating settings but using the same neck locations that has been identified in each program. This experiment shows the neck location is independent of the input arguments. 

\begin{table}[!htb]
\centering
\caption{Neck Miner Results. The second column indicates if there are multiple neck locations to select from. The third column indicates whether the identified neck location is inside the \texttt{main} function}
\scalebox{0.9}{
\begin{tabular}{|c|c|c|}
\hline
\textbf{Program} & \textbf{Multiple Neck Locations} & \textbf{Inside \texttt{main}} \\ \hline
basename 8.32    & \xmark  & \cmark\\ \hline
basenc 8.32      & \xmark  & \cmark\\ \hline
comm 8.32        & \xmark  & \cmark\\ \hline
date 8.32        & \cmark  & \cmark\\ \hline
du 8.32          & \xmark  & \cmark\\ \hline
echo 8.32        & \xmark  & \cmark\\ \hline
fmt 8.32         & \xmark  & \cmark\\ \hline
fold 8.32        & \xmark  & \cmark\\ \hline
head 8.32        & \xmark  & \cmark\\ \hline
id 8.32          & \cmark  & \cmark\\ \hline
kill 8.32        & \xmark  & \cmark\\ \hline
realpath 8.32    & \xmark  & \cmark\\ \hline
sort 8.32        & \xmark  & \cmark\\ \hline
uniq 8.32        & \xmark  & \cmark\\ \hline
wc 8.32          & \cmark  & \cmark\\ \hline
chown-8.2        & \xmark  & \cmark\\ \hline
date-8.21        & \cmark  & \cmark\\ \hline
rm-8.4           & \xmark  & \cmark\\ \hline
sort-8.16        & \xmark  & \cmark\\ \hline
uniq-8.16        & \xmark  & \cmark\\ \hline
gzip-1.2.4       & \cmark  & \xmark\\ \hline
tcpdump-4.10.0   & \xmark  & \cmark\\ \hline
objdump-2.33     & \xmark  & \cmark\\ \hline
readelf-2.33     & \cmark  & \xmark \\ \hline
diff-2.8         & \xmark  & \cmark\\ \hline
Nginx-1.19.0     & \cmark  & \xmark\\ \hline
\end{tabular}
}
\label{tab:neckMinerResults}
\end{table}

\begin{table}[!htb]
\centering
\caption{\rev{Debloating subset apps from Benchmark\_1 based on various input arguments using the same neck location identified in each program}}
\small
\scalebox{0.89}{
\begin{tabular}{|p{0.5cm}|p{1.3cm}|p{2.4cm}|p{0.9cm}|p{0.8cm}|p{0.95cm}|}
\hline
\multirow{2}{*}{App}  & \multirow{2}{*}{\shortstack{Supplied\\ Inputs}} & \multirow{2}{*}{\shortstack{Required\\ Functionality}}                     & \multicolumn{3}{l|}{Reduction After Debloating} \\\cline{4-6} 
                      &  &  & \#Func. & \shortstack{Binary \\ Size} & \shortstack{Total\\ Gadgets}   \\ \hline
\multirow{2}{*}{du}   & -b  & shows number of bytes & 23\% & 15\%  & 46\% \\ \cline{2-6} 
                      & -b --time                        & shows the time of the last modification and number of bytes & 22\%  & 14\% & 45\% \\ \hline
\multirow{3}{*}{sort} & -c     & check if the file given is already sorted or not & 34\%          & 28\%         & 54\%                     \\ \cline{2-6} 
                      & -n & sort a file numerically & 31\%  & 25\%         & 51\% \\ \cline{2-6} 
                      & -un    & sort a file numerically and remove duplicate     & 31\%          & 25\%         & 51\%                     \\ \hline
\multirow{4}{*}{wc} & -c & character count & 42\% & 21\% & 41\% \\ \cline{2-6} 
                      & -w & word count & 42\% & 21\% & 41\% \\\cline{2-6} 
                      & -lc & line and character count & 43\% & 22\% & 42\% \\ \cline{2-6} 
                      & -wc & word and character count & 42\% & 21\% & 42\% \\\hline
\end{tabular}}
\label{tab:capability}
\end{table}

\section{Code Reduction Comparing with other tools}\label{sec:additionalResultsComparison}
This section provides a detailed code reduction comparison with two more debloating approaches.  
\begin{itemize}
    \item Debugger-guided Manual debloating. We developed a simple but systematic protocol to perform debloating manually, which we state as Algorithm~\ref{alg:manualAlg}. The goal of this manual approach is to create an approximation for the maximum level of reduction that can be achieved by an average developer.
    \item Nibbler~\cite{Nibbler2}. state-of-the-art tool for debloating binary code. It does not generate specialized apps, it rather focuses only on reducing the size of shared libraries. 
\end{itemize}

\begin{algorithm}[!htb]
\caption{Debugger-guided Manual Debloating Protocol}
\scriptsize
\label{alg:manualAlg}
\DontPrintSemicolon
\KwIn{App $A$, Input $I$}
\KwOut{App $A\textprime$ }
  $varsToRemove \leftarrow \emptyset$\;\label{alg:l:varsToRemove}
  $funcToRemove \leftarrow \emptyset$\;\label{alg:l:funcToRemove}
  $Executed \leftarrow$ the set of statements that GDB reports are executed, given input $I$ \; \label{alg:l:gdb}
    \SetKwProg{Def}{def}{:}{}
      $A\textprime \leftarrow A$ \;
      \Repeat{no more removals of statements}{
          \For{$stmt \in A\textprime$}{ \label{alg:l:loop1-b}
            \uIf{$stmt \not\in Executed$}{ \label{alg:l:execLst}
               remove stmt from $A\textprime$ \;
               $varsToRemove \leftarrow varsToRemove \cup \{\textit{stmt(vars)}\}$ \;
               \uIf{$stmt$ is a call site}{
                    $funcToRemove \leftarrow funcToRemove \cup \{\textit{func}\}$ \; \label{alg:l:loop1-e}
               }
            }
      \While{$varsToRemove$ $\neq \emptyset$ $ \land $ $funcToRemove$ $\neq  \emptyset$}{ \label{alg:l:loop2-b}
            \For{$func \in funcToRemove$}{
            Remove $func$ from funcToRemove \;
            \uIf{no occurrence of $func$ exists}{\label{alg:l:occrFunc}
                Remove func from A\textprime \;
            }
          }
          \For{$var \in varsToRemove$}{\label{alg:l:occrVar}
            Remove $var$ from varToRemove \;
            \uIf{no occurrence of $var$ exists}{
                Remove var from A\textprime \;\label{alg:l:loop2-e}
            }
          }
      }
      \uIf{$A\textprime$ does not build correctly}{\label{alg:l:build-b}
        put back $stmt$ \;
        undo $var$ and $func$ removal from A\textprime\; \label{alg:l:undo}
      }\label{alg:l:build-e}
      }
     }
\end{algorithm}

\begin{figure}[!htb]
    \centering
    \includegraphics[width=1\columnwidth]{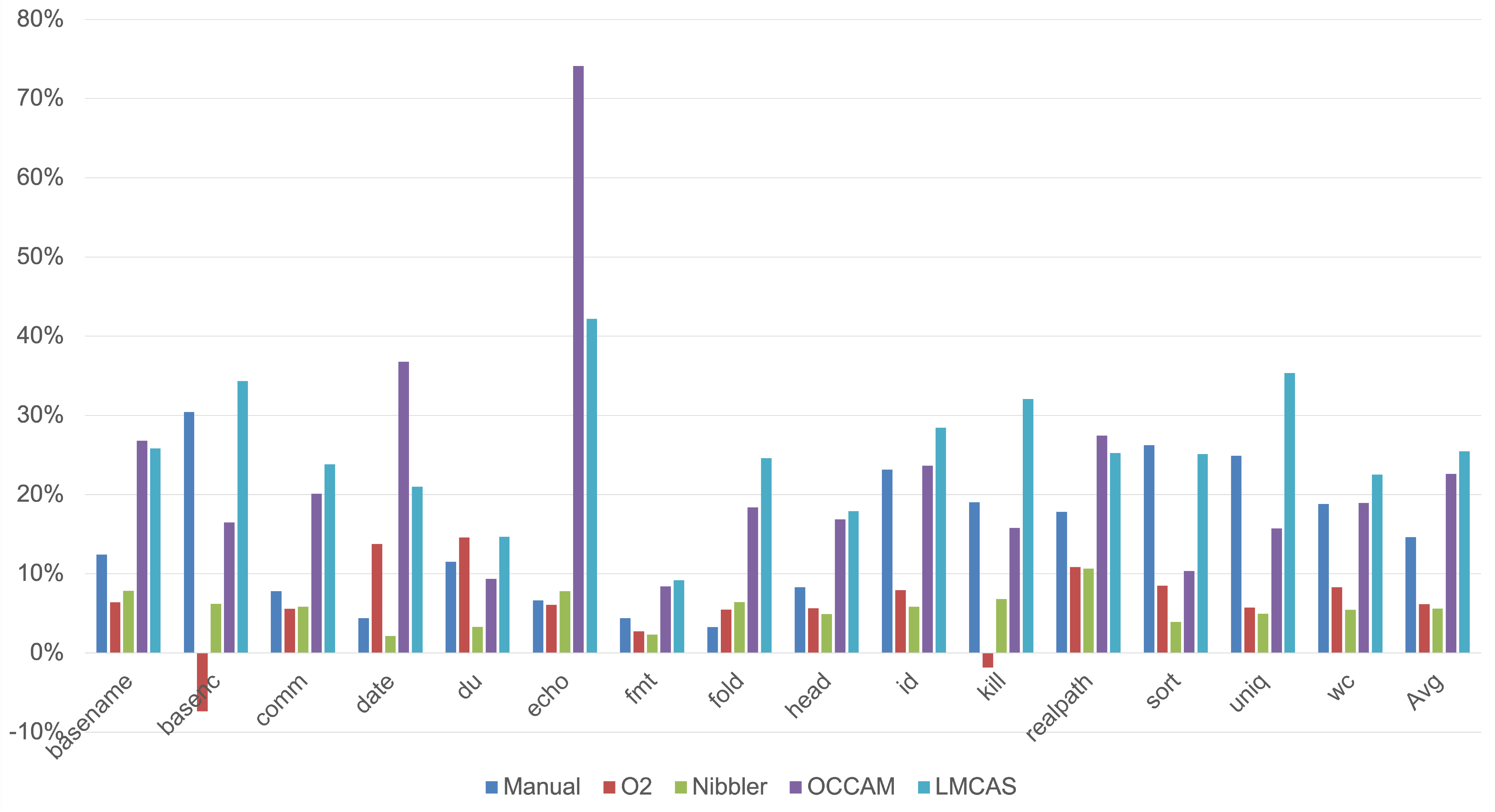}
    \caption{Binary size reduction achieved through manual debloating, baseline, Nibbler, OCCAM, and \lmcas. (Higher numbers are better.)}
    \label{fig:avgReduction}
\end{figure}

We used \textit{Benchmark\_1} to compare the performance of \lmcas against manual debloating, baseline, Nibbler, and OCCAM.
Figure~\ref{fig:avgReduction} shows the comparison results based on the reduction in the binary size that each tool achieved for each app in \textit{Benchmark\_1}. For computing the binary-size metric, we compiled all debloated apps with \texttt{gcc -O2}, and ran \texttt{size}.


\section{\lmcas Running Time}\label{apndx:runTimeLMCAS}
\begin{figure}[!htb]
    \centering
    \includegraphics[width=1\columnwidth]{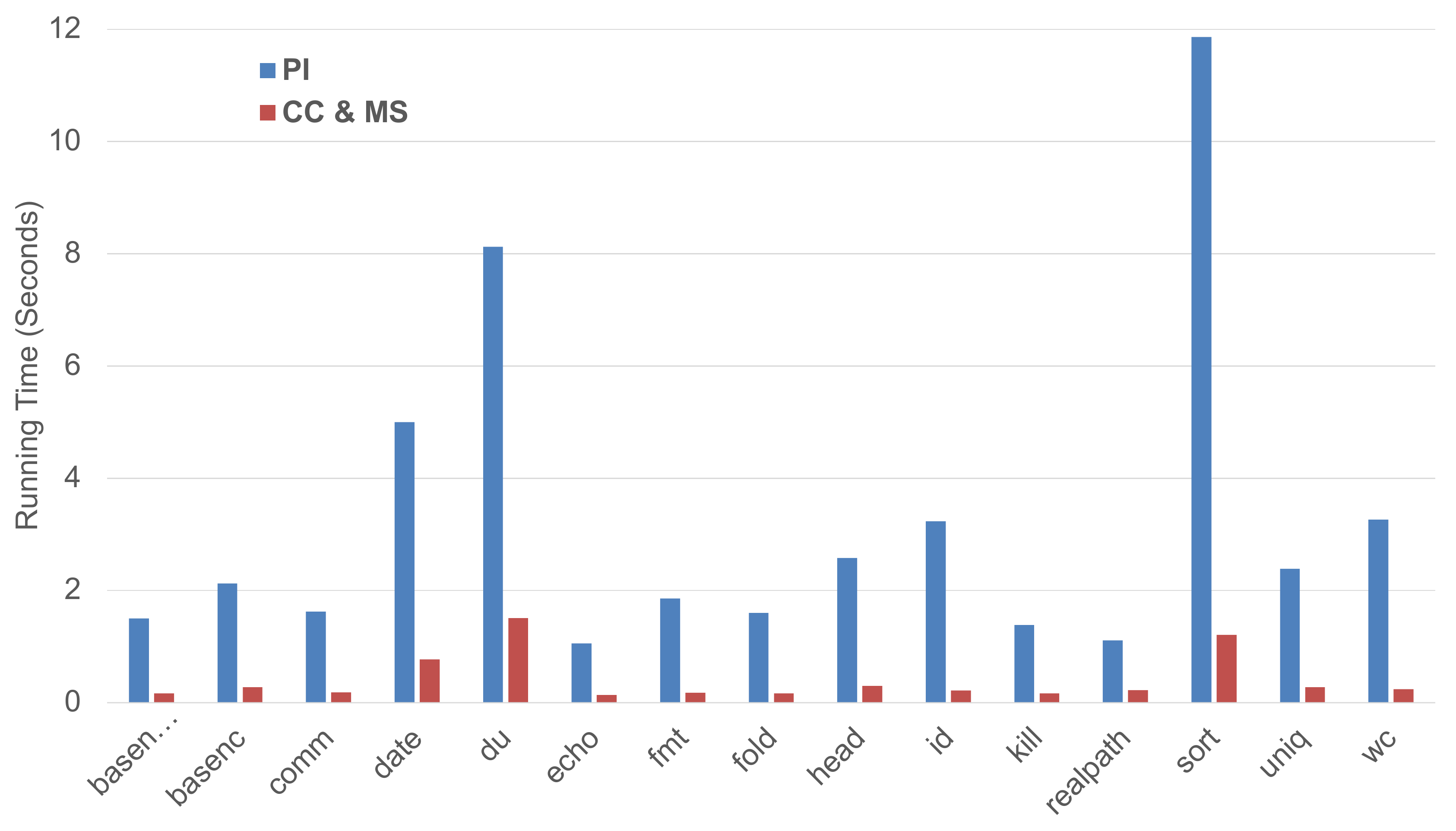}
    \caption{The time required for partial interpretation (PI) and the partial-evaluation steps (Constant Conversion and Multi-stage Simplifications) for \textit{Benchmark\_1}.
    \Omit{\twr{Please change the label on the y-axis to "Running Time", and change the
    key from "CC \& MS" to "Partial Evaluation" to be consistent with the caption. We need \underline{all} labeling and terminology to be consistent \underline{everywhere}.}}}
    \label{fig:analysisTime}
\end{figure}

We measured the running time of \lmcas. Figure~\ref{fig:analysisTime} shows the breakdown of running time, for \textit{Benchmark\_1}, between (i) partial interpretation, and (ii) Constant Conversion (CS) plus Multi-stage Simplification (MS).

The average total running time is $3.65$ seconds;
the maximum total running time is $13.08$ seconds for analyzing \texttt{sort};
and the lowest total analysis time is $1.19$ seconds for analyzing \texttt{basename}.
Notably, the time for Constant Conversion and Multi-stage Simplification is low:
on average, the time for constant conversion and multi-stage simplification is $0.4$ seconds, while the average time for Partial Interpretation is $3.25$ seconds.  

\rev{
}

\section{CFI Experiment} \label{apndx:cfi}

\begin{table}[!htb]
\centering
\caption{Total unique gadgets count for original binaries, debloated binaries using \lmcas}
\small
\scalebox{0.92}{
\begin{tabular}{|c|c|c|c|c|}
\hline
\multirow{2}{*}{\textbf{Program}} & \multicolumn{4}{c|}{\textbf{Total   ROP Count}}    \\ \cline{2-5}
         & \multicolumn{1}{c|}{\textbf{Original}} & \textbf{LLVM-CFI} & \textbf{LMCAS} & \textbf{\shortstack{LMCAS\\+LLVM-CFI}}\\ \hline
basename & 1794         & 964               & 841       &  261   \\
basenc   & 3063         & 1805              & 1309      &  793  \\
comm     & 2095         & 1145              & 964       &  794   \\
date     & 12119        & 3654              & 3381      &  1592   \\
du       & 15094        & 7874              & 8503      &   5873  \\
echo     & 1835         & 446               & 876       &   442  \\
fmt      & 2496         & 1403              & 1230      &   1158  \\
fold     & 2094         & 1168              & 1015      &  769   \\
head     & 2671         & 1398              & 1366      &   932  \\
id       & 2514         & 1214              & 1183      &   801  \\
kill     & 1924         & 1147              & 919       &   1054  \\
realpath & 3073         & 1610              & 1664      &   1658  \\
sort     & 7558         & 4804              & 4185      &  3699   \\
uniq     & 2516         & 1280              & 1121      &   776  \\
wc       & 2225         & 1611              & 1320      &   973  \\
objdump  & 115587       & 103241            & 107985    &   80156  \\
readelf  & 58186        & 50512             & 56519     &  45320   \\
tcpdump  & 82682        & 53205             & 67417     &  50809   \\
Chown    & 2890         & 2280              & 2529      &   1998  \\
rm       & 3068         & 2316              & 2579      &  2083  \\ \hline
\end{tabular}}
\label{tab:cfiAnalysis}
\end{table}

